\documentclass[twocolumn]{aastex63}
\usepackage{amsmath}

\usepackage{graphicx}
\usepackage{amsmath}
\usepackage{mathtools}
\usepackage{kantlipsum}
\usepackage{xparse}
\usepackage{enumerate}
\usepackage[version=3]{mhchem}
\usepackage{amsmath,array}
\usepackage[T1]{fontenc}
\usepackage{booktabs}

\newcommand*\VF[1]{\mathbf{#1}}
\usepackage{sidecap}
\usepackage{tabularx}

\received{Received}
\revised{Revised}
\accepted{Accepted}
\submitjournal{ApJ}

\graphicspath{{./}{figures/}}

\begin{document}

\title{Modeling the high-resolution emission spectra of clear and cloudy non-transiting hot Jupiters}

\correspondingauthor{Isaac Malsky}
\email{imalsky@umich.edu}

\author[0000-0003-0217-3880]{Isaac Malsky}
\affil{Department of Astronomy and Astrophysics, University of Michigan, Ann Arbor, MI, 48109, USA}

\author[0000-0003-3963-9672]{Emily Rauscher}
\affil{Department of Astronomy and Astrophysics, University of Michigan, Ann Arbor, MI, 48109, USA}

\author[0000-0002-1337-9051]{Eliza M.-R.\ Kempton}
\affil{Department of Astronomy, University of Maryland, College Park, MD 20742, USA}

\author[0000-0001-8206-2165]{Michael Roman}
\affil{School of Physics and Astronomy, University of Leicester, University Road, Leicester LE1 7RH, UK}

\author[0000-0003-3840-7490]{Deryl Long}
\affil{Department of Astronomy and Astrophysics, University of Michigan, Ann Arbor, MI, 48109, USA}
\affil{Department of Astronomy and Astrophysics, University of Virginia, Charlottesville, VA, 22904, USA}

\author[0000-0001-5737-1687]{Caleb K. Harada}
\altaffiliation{NSF Graduate Research Fellow}
\affil{Department of Astronomy, University of California Berkeley, Berkeley, CA 94720, USA}
\affil{Department of Astronomy, University of Maryland, College Park, MD 20742, USA}

\begin{abstract}
The advent of high-resolution spectroscopy as a method for exoplanet atmospheric characterization has expanded our capability to study non-transiting planets, increasing the number of planets accessible for observation. Many of the most favorable targets for atmospheric characterization are hot Jupiters, where we expect large spatial variation in physical conditions such as temperature, wind speed, and cloud coverage, making viewing geometry important. Three-dimensional models have generally simulated observational properties of hot Jupiters assuming edge-on viewing, which neglects planets without near edge-on orbits. As the first investigation of how orbital inclination manifests in high-resolution emission spectra, we use a General Circulation Model to simulate the atmospheric structure of Upsilon Andromedae b, a non-transiting hot Jupiter. In order to accurately capture scattering from clouds, we implement a generalized two‐stream radiative transfer routine for inhomogeneous multiple scattering atmospheres. We compare models with and without clouds, as cloud coverage intensifies spatial variations. Cloud coverage increases both the net Doppler shifts and the variation of the continuum flux amplitude over the course of the planet’s orbit. As orbital inclination decreases, four key features also decrease in both the clear and cloudy models: 1) the average continuum flux level, 2) the amplitude of the variation in continuum with orbital phase, 3) net Doppler shifts of spectral lines, and 4) Doppler broadening in the spectra. Models capable of treating inhomogeneous cloud coverage and different viewing geometries are critical in understanding high-resolution emission spectra, enabling an additional avenue to investigate these extreme atmospheres.
\end{abstract}
\keywords{planets and satellites: atmospheres}

\section{Introduction}\label{sec:Introduction}
Beginning with 51 Pegasi b, exoplanet surveys have discovered a number of so-called hot Jupiter planets unlike anything in our solar system (\citealt{1995Natur.378..355M, 2016A&A...587A..49O}). These planets are massive gas giants (akin to Jupiter) and have orbital semi-major axes less than 0.05 au. The high temperatures and close orbits of these planets have made them ideal candidates for atmospheric characterization \citep{doi:10.1146/annurev-astro-081309-130837}. By observing the molecular spectral features and temperature structures of hot Jupiters, we can gain insight into a realm of exoplanet and atmospheric physics with no solar system analogue.

The atmospheres of hot Jupiters are expected to have extreme spatial variations in their temperature structures \citep{2002A&A...385..166S}. Tidal torques and viscous dissipation are expected to lock hot Jupiters into synchronous rotation in relatively short timescales, leading to planets with permanent daysides and constant nightsides \citep{1996ApJ...459L..35G, 1996ApJ...470.1187R}. Across the population of hot Jupiters, observations show dayside temperatures exceeding 2500\,K \citep[e.g.,][]{10.1093/mnras/stv470, 2020A&A...639A..36B, 2020AJ....159..137G}, and nightside temperatures of approximately 1100\,K \citep[e.g.,][]{2019NatAs...3.1092K, 2019AJ....158..166B}. These large temperature differences cannot exist apart from a global circulation pattern that is working (somewhat ineffectually) to reduce the day-night temperature gradient. The diversity and expected complexity of hot Jupiters mean 3D simulations of the atmospheric conditions are critical to interpreting observational results.

Computational simulations are a critical tool for understanding the properties of hot Jupiters, as they connect observations and theoretical models. Significant work has been done creating General Circulation Models (GCMs) to simulate the atmospheric dynamics of hot Jupiters \citep[e.g.,][]{2009ApJ...699..564S, 2010ApJ...710.1395D, RauscherMenou2010, 10.1111/j.1365-2966.2011.18315.x, 2014A&A...561A...1M}. There are standard common features between models: transonic winds, including a broad equatorial jet that advects hot gas eastward away from the substellar point, and large day and nightside temperature differences.

The first use of high-resolution spectroscopy to characterize an exoplanet atmosphere detected hints of atmospheric winds via a slight Doppler shift in HD 209458b \citep{2010Natur.465.1049S}. This technique, which uses the orbital motion of the planet to separate its spectrum from the stellar and telluric components, has been shown to be sensitive to winds and rotational velocities of exoplanets \citep{2018arXiv180604617B}. Furthermore, these spectral line shifts can be used as a probe for observed exoplanet atmospheres, revealing their 3D structure when interpreted in comparison against spectra generated from GCM simulations \citep[e.g.,][]{Flowers_2019, Beltz}. Recent simulations of planets with edge-on transiting orbits have shown that the winds, rotation, and thermal structure of hot Jupiters produce broadening and net Doppler shifts on the order of 1\,km/s in their disk integrated high-resolution thermal emission and transmission  spectra \citep{2017ApJ...851...84Z} and that the 3D thermal structure of hot Jupiters is observable in high-resolution emission spectroscopy measurements \citep{Flowers_2019, Beltz}. The addition of clouds can further complicate the influence of transonic winds and planet rotation on the net Doppler shift, as shown by \cite{harada2021}.

Cloud coverage can result in significant atmospheric absorption and scattering and can compound the heterogeneous nature of hot Jupiters \citep{doi:10.1146/annurev-earth-060614-105146}. Several observational results, including evidence of scattering or gray opacity in transmission spectra and weak water absorption features in near IR emission spectra, point to the presence of aerosols in exoplanet atmospheres \citep{Crossfield2013, 2014Natur.513..526F, 2014ApJ...794..155K, 2014Natur.505...66K,helling2019exoplanet}. Additionally, phase curve observations provide evidence for spatially inhomogeneous clouds \citep{parm2018}. In the infrared, thermal emission phase curve observations have shown lower nightside temperatures than would be expected from cloud-free models \citep[e.g.,][]{2014Sci...346..838S}. In the optical, reflected light phase curve observations have shown spatial structure in the albedo patterns \citep{Demory2013}. One proposed explanation of this is that clouds condense on the colder nightside and push the photosphere to higher altitudes, decreasing the nightside brightness temperature. Then, as the eastward wind pattern advects cold air and clouds from the nightside around to the dayside, clouds may enhance reflection to the west of the substellar point before the gas heats and the clouds dissipate \citep{Kataria_2015, Stevenson_2017, parmentier2021cloudy, roman2021clouds}.

Recent works have included clouds in 3D models of hot Jupiter atmospheres \citep{parmientier2016,lee+2016,lee+2017,lines+2018,lines2019overcast,2019ApJ...872....1R,parmentier2021cloudy, roman2021clouds}. The presence of aerosols uniformly leads to decreased nightside fluxes, altered day-night temperature contrasts, and hot-spot offsets, especially when the cloud radiative feedback is included \citep{lines2019overcast, 2019ApJ...872....1R}. Comparing these models with detailed observations will help us to constrain the nature of the atmospheric aerosols.

\begin{figure*}[!htb]
\begin{center}
\includegraphics[width=\linewidth]{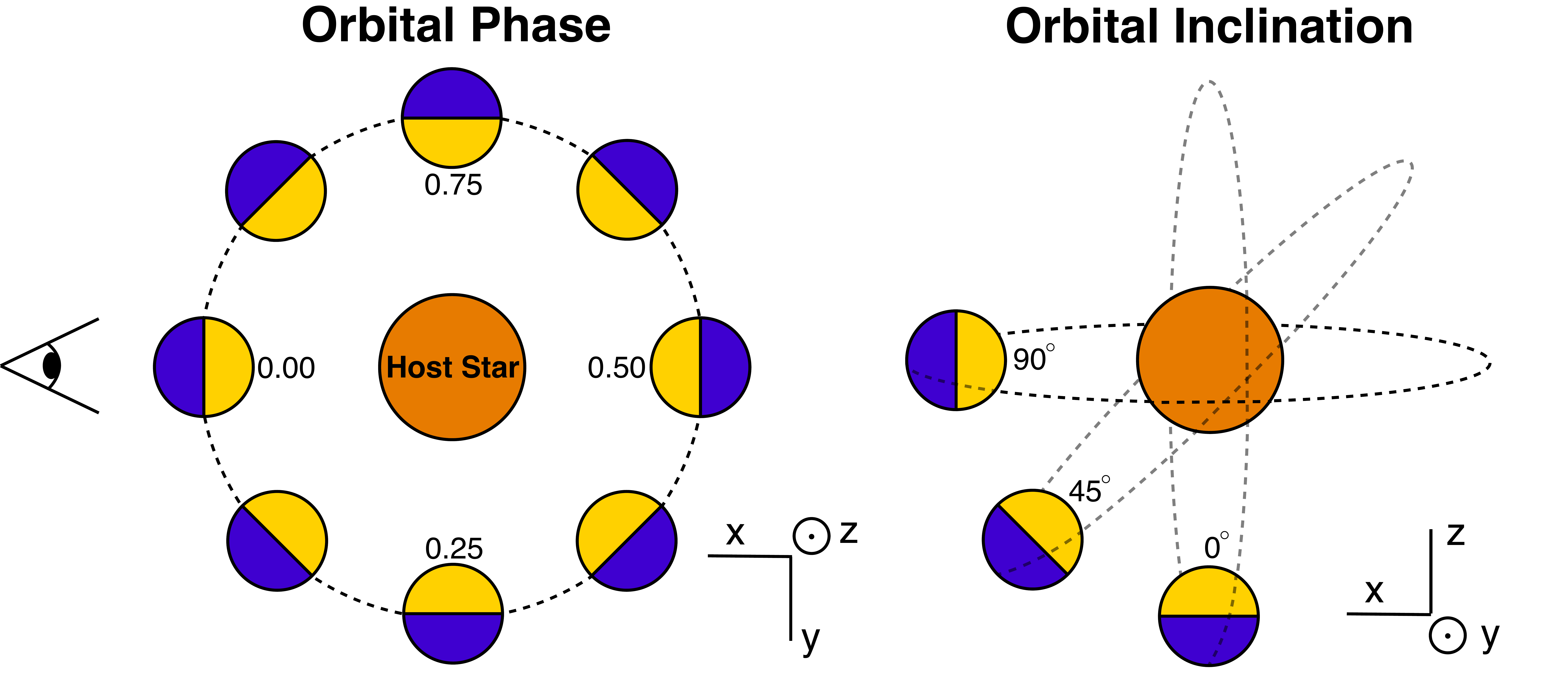}
\caption{An illustration of the geometry of the observer, planet, and host star system that we consider when simulating emission spectra. Left, a planet with a 90$^{\circ}$ inclination, at different phases as indicated. We rotate the 3D GCM model for a number of phases. For this orbital motion the angular momentum vector points out of the page along the z axis. Because the planet is synchronously rotating so that it maintains a constant dayside, the portion of the planet's dayside hemisphere visible to an Earth observer changes throughout its orbit. A phase of 0$^{\circ}$ corresponds to a nightside emission spectrum. Right, an example of the model at three different orbital inclinations. An orbital inclination of 90$^{\circ}$ corresponds to edge-on, while 0$^{\circ}$ corresponds to pole on. Due to the geometry of the orbit, a planet with an inclination of 0$^{\circ}$ will present the same hemisphere to the observer no matter the phase.} 
\label{fig:drawing}
\end{center}
\end{figure*}

To date, most emission spectra simulated from 3D models (and all high-resolution spectra) are calculated for the equatorial viewing geometry applicable to transiting systems. Because the orbital planes of exoplanets are random relative to us, a relatively small fraction of hot Jupiters transit in front of their host star \citep{Wright2012}. See Figure~\ref{fig:drawing} for an illustration of the system geometry. Expanding modeling capabilities to characterize planets with non-edge-on inclinations greatly increases the flexibility of GCMs to interpret high-resolution spectroscopy results. High-resolution emission spectroscopy observations only rely on a planet's orbital motion and so can be applied to both transiting and non-transiting planets. We explore how the 3D structure of hot Jupiter atmospheres, as expressed in high-resolution spectra, varies with changing orbital phase and inclination.

In order to study the effects of orbital inclination on the high-resolution emission spectra of clear and cloudy hot Jupiters, we apply the 3D GCM from \cite{rauscher2012general} with radiatively active clouds as modeled in \cite{2019ApJ...872....1R} to Upsilon Andromedae b, a non-transiting hot Jupiter that has been extensively characterized \citep[e.g.,][]{2001A&A...367..943J, 2009IAUS..253..239H, Piskorz_2017}. A fairly standard hot Jupiter, with an orbital period of $\sim$4.6 days \citep{Butler_2006} and a mass 1.7 times that of Jupiter \citep{Piskorz_2017}, its orbital inclination of 24 degrees \citep{Piskorz_2017} means that we view more of its poles than its equator. Other properties of Upsilon Andromedae b are shown in Table \ref{tab:params}.

\begin{deluxetable*}{lllll}\label{tab:params}
{\tablehead{\multicolumn{5}{c}{\textit{System Parameters for Upsilon Andromedae b}}}}
\startdata
\multicolumn{1}{c}{Parameter} & \multicolumn{2}{c}{Value}  & \multicolumn{1}{c}{Units} & \multicolumn{1}{c}{Reference}\\
\hline
Orbital Period            &\multicolumn{2}{c}{4.617}                 & days             & \cite{Butler_2006}\\
Semi-major Axis                     &\multicolumn{2}{c}{0.0595}                & au               & \cite{Butler_2006}\\
Orbital Inclination        &\multicolumn{2}{c}{24}                    & degrees          & \cite{Piskorz_2017}\\
Stellar Effective Temperature       &\multicolumn{2}{c}{6212}                  & K                & \cite{santos2013} \\
Stellar Radius                      &\multicolumn{2}{c}{1.0296$\times$10$^9$}  & m                & \cite{2009ApJ...694.1085V}\\
\hline
\multicolumn{5}{c}{\textit{Shared Model Parameters}}\\
\hline
\multicolumn{1}{c}{Parameter} & \multicolumn{2}{c}{Value}  & \multicolumn{1}{c}{Units} & \multicolumn{1}{c}{Reference}\\
\hline
Planetary Rotation Rate                       &\multicolumn{2}{c}{1.58$\times$10$^{-5}$} & radians s$^{-1}$ & \\
Infrared absorption coefficient $\kappa_{\mathrm{IR}}$ &\multicolumn{2}{c}{1.08$\times$10$^{-2}$} & cm$^2$ g$^{-1}$  & \cite{RomanRauscher2017}\\
Optical absorption coefficient $\kappa_{\mathrm{VIS}}$ &\multicolumn{2}{c}{1.57$\times$10$^{-3}$} & cm$^2$ g$^{-1}$  & \cite{RomanRauscher2017}\\
\hline
\multicolumn{5}{c}{\textit{Gravity Assumptions}}\\
\hline
\multicolumn{1}{c}{Parameter} & \multicolumn{1}{c}{Low Gravity}  & \multicolumn{1}{c}{High Gravity}  & \multicolumn{1}{c}{Units} & \multicolumn{1}{c}{Reference}\\
\hline
Gravitational Acceleration                      & 12.9324                       & 24.7935             & m  s$^{-2}$     & \cite{2010ApJ...723.1436C,Piskorz_2017}\\
Radius                                          & 1.287$\times$10$^8$           & 9.294$\times$10$^7$ & m               & \cite{2010ApJ...723.1436C,Piskorz_2017}\\
IR photospheric pres ($\tau$=2/3)               & 80                            & 153                 & mbar                &
\enddata
\caption{The planetary parameters that we adopt for the simulations of Upsilon Andromedae b. Here we present three GCMs: low gravity with no clouds, high gravity with no clouds, and low gravity with clouds. The IR photospheric pressure is the $\tau$=2/3 level for a clear atmosphere. The double-gray absorption coefficients were chosen to match those used in \cite{2019ApJ...872....1R} and \cite{harada2021} for consistency and comparison between models. Last, the rotation rate was determined from the orbital period in \cite{Butler_2006} under the assumption that the planet is synchronously rotating.}
\end{deluxetable*}

Spitzer phase curve observations of this planet showed a large amplitude and large hotspot offset \citep{2006Sci...314..623H,2010ApJ...723.1436C}.
These results are evidence of an almost 1000\ K, day-night temperature difference and strong winds to significantly displace the brightest point on the planet away from the substellar point, reaffirming the need for any model of Upsilon Andromedae b to be fully three dimensional. We create simulated spectra at this planet's measured orbital inclination, but also use this as a template to study the signature of 3D effects in high-resolution spectra at a range of inclinations. By simulating a range of orbital inclinations and phases we show how spatial variations in the atmosphere are manifest in simulated high-resolution emission spectra.

We present our methodology for modeling the atmospheric circulation on hot Jupiters and results as follows: in \S~\ref{sec:Methods}, we detail our 3-D GCM and the cloud treatment we use, describe our methods for remapping the GCM for arbitrary phases and inclinations, and how we generate high-resolution emission spectra for each. Here we also describe our upgrade to the radiative transfer post-processing, with further details given in the Appendix. In \S~\ref{sec:Results}, we present results for the clear and cloudy models of Upsilon Andromedae b, including the simulated emission spectra for this planet. Next, we explore how viewing a planet at different inclinations changes the resulting emission spectra, including the Doppler signatures they contain. Finally, a discussion of the results and the conclusions are in \S~\ref{sec:conclusions}.

\section{Methods}\label{sec:Methods}
In order to connect predicted inhomogeneous physical conditions on a hot Jupiter with observable features in high-resolution emission spectra, we run 3D models of Upsilon Andromedae b, which is a planet of high observational interest and also serves here as an example with which to study non-transiting planets more generally and use them to simulate spectra for a variety of orbital phases and viewing inclinations. First, we model Upsilon Andromedae b following the methods of \cite{2019ApJ...872....1R}. We use a GCM to simulate clear and cloudy atmospheres for cases assuming high and low estimates for the planet's gravity. High-resolution characterization can constrain a non-transiting planet's mass by measuring the orbital inclination and breaking the $m_p \sin i$ degeneracy, but with no transit we cannot directly constrain the planet's radius.

In order to calculate simulated spectra from these 3-D models, we remap the outputs to the viewing geometry corresponding to the inclination and phase of observation (as required by the radiative transfer post-processing code). We begin by calculating spectra for the known orbital inclination of Upsilon Andromedae b, before expanding to consider spectra as observed over a range of possible inclinations to generally explore the impact of orbital inclination on high-resolution observations. For all spectra, we calculate versions with and without Doppler shifts. Last, we cross correlate the spectra with their un-Doppler-shifted counterparts to assess the Doppler contributions of planet rotation and winds.

\subsection{3D General Circulation Model}\label{subsec:3d}
We use a GCM previously developed and described in \cite{RauscherMenou2010, rauscher2012general} and \cite{RomanRauscher2017,2019ApJ...872....1R} to compute atmospheric winds and temperatures given flux boundary conditions and relevant planetary parameters. The GCM solves the two-stream radiative transfer equations and the ``primitive'' equations of meteorology to calculate atmospheric radiative heating rates for both clear and cloudy conditions using the double-gray approximation in which infrared and optical frequencies are treated separately \citep{1989JGR....9416287T}. The double-gray assumption uses these two parameters ($\kappa_{\rm{IR}}$ and $\kappa_{\rm{VIS}}$) to capture the overall heating and cooling rates within the atmosphere, resulting in temperature-pressure profiles that can be solved for analytically \citep{Guillot2010}.

We chose the same gaseous infrared and visible absorption coefficients as \cite{RomanRauscher2017}. We find that this set of absorption coefficients produces reasonable temperature-pressure profiles for planets at a variety of irradiation temperatures, in comparison to results from more detailed 1D models, and so adopt them again in this work in order to allow for easier comparison with our models of other planets. Note that this choice is also assuming a non-inverted temperature profile on the planet.

For our current work, we utilize the simple cloud modeling scheme of \cite{2019ApJ...872....1R}, which treats clouds as temperature-dependent scatterers and absorbers that shape the atmosphere through radiative effects and form or dissipate as the temperature structure evolves. Although this scheme does not attempt to model microphysical processes and self-consistent chemistry, it has been shown to favorably agree with basic cloud distributions and radiative effects found in more rigorous modeling \citep{lee+2016, lee+2017, lines+2018, lines2019overcast}.

Our GCM simulations parameterize the planet using the values given in Table~\ref{tab:params}, and follow the same procedures as in \cite{2019ApJ...872....1R}. In all cases, the planet is assumed to be synchronously rotating, with an orbital period exactly equal to the rotational period, resulting in a permanent dayside and permanent nightside. We model the atmosphere from 10$^{-4}$ to 10$^{2}$ bar, using 50 vertical layers evenly spaced in log pressure, with a horizontal spectral resolution of T31, which corresponds to a spatial grid of 48 latitudes by 96 longitudes. Simulations are run for 2000 orbital periods at 4800 time-steps per period, which we find sufficient for results to converge.

Upsilon Andromedae b does not transit its host star, and therefore we have no direct measurement of its radius. To account for this, we ran simulations with two different surface gravity values, chosen to span the radius parameter space allowed by the day-night flux variation seen in \cite{2010ApJ...723.1436C} under the inclination constraint from \cite{Piskorz_2017}. The low-gravity model has a surface gravity of 12.9 m/s$^2$ and a radius of 1.8 R$_J$ and the high-gravity model had a surface gravity of 24.8 m/s$^2$ and a radius of 1.3 R$_J$. These values serve as lower and upper bound estimates for the radius and surface gravity of Upsilon Andromedae b. While these parameters have been chosen to match the specific planet Upsilon Andromedae b, they are fairly typical hot Jupiter values and so are appropriate for our more general exploration of orbital inclinations later in this paper.

We model Upsilon Andromedae b with and without clouds in order to study the effects of inhomogeneous aerosols. In cloudy cases, clouds will form when the atmospheric temperature profile drops below the condensation temperature for a particular cloud species. We include four species: MnS (alabandite), Al$_2$O$_3$ (corundum), Fe (iron), and MgSiO$_3$ (enstatite), following \cite{2019ApJ...872....1R} and \citet{harada2021} for consistency. For comparison, we create similar models without clouds, which we call the `clear' models. Not only does this difference affect the final emission spectra, but it also impacts the atmospheric dynamics and temperature structure within the GCM because of ongoing cloud feedback throughout the model evolution \citep{2019ApJ...872....1R}. Our chosen values for the cloud properties within the GCM are shown in Table~\ref{tab:params2}.

\begin{deluxetable*}{lllll}[t]\label{tab:params2}
\tablehead{ \colhead{Parameter} & \colhead{MgSiO$_3$} & \colhead{Fe} & \colhead{Al$_2$O$_3$} & \colhead{MnS}}
\startdata
Molecular weight, $\mu_g$ [g mol$^{-1}$] & 100.4 & 55.8 & 102.0 & 87.0 \\
Mole fraction, $\chi_g$ & 3.26$\times$10$^{-5}$ & 2.94$\times$10$^{-5}$ & 2.77$\times$10$^{-6}$ &  3.11$\times$10$^{-7}$\\
Particle Density, $\rho$ [g cm$^{-3}$] & 3.2 & 7.9 & 4.0 & 4.0\\
Refractive index, n & 1.5 + (4.0$\times$10$^{-4}$)$i$ & 4.1 + 8.3$i$ & 1.6 + (2.0$\times$10$^{-2}$)$i$ & 2.6 + (1.0$\times$10$^{-9}$)$i$\\
Extinction efficiency ($\lambda=2.3\,\mu$m), $Q_{2.3\mu\text{m}}$ & 0.07 & 1.25 & 0.12 & 0.56 \\
Extinction efficiency ($\lambda=5\,\mu$m), $Q_{5.0\mu\text{m}}$ & 0.01 & 0.16 & 0.02 & 0.02 \\
Optical depth per bar, $\tau/\Delta P$ & 29.0 & 105.1 & 3.4 & 1.5 \\
\enddata
\caption{The aerosol properties for the clouds in our GCM. Optical depths per bar are from \cite{2019ApJ...872....1R}. Scattering parameters and extinction efficiencies were calculated in \cite{harada2021}, with complex refractive indices from \cite{kitzmann}.}
\end{deluxetable*}

We use the cloud locations and vertical extents output from the GCM in our emission spectrum post-processing to model the radiative transfer of light through the atmosphere as described in \S~\ref{subsec:1}. The vertical distribution and optical thickness of the clouds will, in reality, depend on the interplay between vertical mixing and intricate microphysical processes, but to retain the numerical efficiency and modest simplicity of our cloud modeling, we assume a distribution in which clouds are extended vertically above the heights at which they condense. This implicitly assumes the atmosphere is well-mixed and cloud particles can be easily lofted far above the condensation level, which is roughly consistent with results of more rigorous modeling \citep[e.g.\ studies including dynamical tracers such as][]{2013A&A...558A..91P} and rigorous cloud microphysics \citep{lines+2018}. However, while the clouds are expected to extend vertically over multiple pressure scale heights, a majority of the cloud mass is expected be found at the base of the cloud \citep{Powell_2018}.  To roughly mimic this behavior, we assume that cloud can extend up to 40 layers (roughly eight pressure scale heights) above the base of the cloud, temperatures permitting, reaching up to the $\sim$ 0.5 mbar level. Within the cloud, the mass of condensate is set to be proportional to the gas pressure within each layer, thus increasing in mass with depth. Following \cite{2019ApJ...872....1R} and \cite{harada2021}, the cloud mass is limited to 1/10th of the maximum potential mass that would occur if all the vapor condensed. Details on the scattering properties of the clouds are discussed in Section 2.3. 

The assumed vertical extent of the atmospheric clouds influences the magnitude of atmospheric scattering. As shown in \cite{2019ApJ...872....1R}, if the clouds are vertically compact they may be hidden deeper in the atmosphere and therefore have a smaller effect on the emission spectra. This outcome is particularly important near the poles, where the difference in vertical range of the compact vs extended clouds is maximized \citep{2019ApJ...872....1R}. Therefore, for the low inclination models presented here, the results will be further from the clear cases when clouds are more vertically extended. The extended cloud case represents the maximum possible effect of clouds, while the compact cloud parameterization has an impact in between the extended and clear atmosphere results for thermal observables.

\subsection{Varying Inclination and Phase}\label{subsec:gridding}
In order to model the emission spectra of hot Jupiters with non-zero inclinations, we create a routine to re-grid temperature, pressure, and wind vector data for arbitrary planet phases and inclinations. The radiative transfer post-processing code requires a planet coordinate system in which the sub-observer point is at $\phi$=0, $\theta$=0 (latitude and longitude, respectively), and the equator is tangent to the plane of the sky. This is specifically what prompts the coordinate system transform and also explains the geometry that we need to achieve. The post-processing uses ray-tracing in the direction of the observer and accounts for Doppler shifts from local wind and rotational velocities. Figure \ref{fig:drawing} shows the system geometry for different phases and inclinations. The planet is tidally locked, and the phase and inclination determines which portion of the planet is visible to an observer.

We remap the GCM output from constant pressure levels to 250 levels of constant altitude using vertical hydrostatic equilibrium, consistent with the GCM framework. The choice of 250 layers is elaborated on in detail in the Appendix. In short, 250 layers reduced the fractional error of our results to below 1\%, well below the expected error inherent in other model assumptions. For each atmospheric grid location there is an associated latitude, longitude, temperature, pressure, altitude, $u$, $v$, and $w$ (east-west, north-south, and up-down) wind components, and various cloud properties. With the remapped grid, we then use a radial basis function to interpolate the temperature, pressure, and wind data onto a regularly spaced grid of 48 latitude and 96 longitude points. Throughout this work, we chose these values to maintain an approximately equal grid resolution with the GCM.

Accounting for changes in phase is simply a matter of changing the longitude grid values. To account for varying inclination however, each structure variable (e.g. temperature, pressure, cloud properties) must be remapped and interpolated. Further, due to their vector nature, it is necessary to treat each component of the wind vectors separately. In order to rotate the 3-D wind components we first calculate their values in a Cartesian grid as,

\begin{align}
  \phantom{i + j + k}
  &\begin{aligned}
    \mathllap{v_x} = u \sin(\theta - \theta_{obs}) + v \cos(\theta - \theta_{obs}) \sin(\phi) \\ - w \cos(\theta - \theta_{obs}) \cos(\phi)
  \end{aligned}
\end{align}

\begin{align}
  \phantom{i + j + k}
  &\begin{aligned}
    \mathllap{v_y} = -u \cos(\theta-\theta_{obs}) \ + \\
    v \sin(\theta-\theta_{obs}) \sin(\phi) + w \sin(\phi)
  \end{aligned}
\end{align}

\begin{align}
  \phantom{i + j + k}
  &\begin{aligned}
  v_z = v \cos(\phi) - w \sin(\theta- \theta_{obs}) \cos(\phi)
  \end{aligned}
\end{align}

\noindent where $\theta_{obs}$ is the sub-observer longitude. We then rotate the planet grid about the y-axis (consistent with the geometry in Figure \ref{fig:drawing}) according to the choice of inclination and convert from Cartesian velocity components back to the locally defined $u$, $v$, and $w$ components as,

\begin{equation}
  \begin{bmatrix}
    v_{x}' \\
    v_{y}' \\
    v_{z}' \\
  \end{bmatrix}
  =
  \begin{bmatrix}
    \cos(\delta) & 0 & \sin(\delta)\\
    0 & 1 & 0 \\
    -\sin(\delta) & 0 & \cos(\delta)\\
  \end{bmatrix}
  \begin{bmatrix}
    v_x\\
    v_y\\
    v_z\\
  \end{bmatrix}
\end{equation}

\begin{align}
  \phantom{i + j + k}
  &\begin{aligned}
    \mathllap{u'} = -v_x' \sin(\theta') + v_y' \cos(\theta'),
  \end{aligned}
\end{align}

\begin{align}
  \phantom{i + j + k}
  &\begin{aligned}
    \mathllap{v'} = \sin(\phi') [-v_x' \cos(\theta') - v_y' \sin(\theta')] \\ + v_z' \cos(\phi'),
  \end{aligned}
\end{align}

\begin{align}
  \phantom{i + j + k}
  &\begin{aligned}
    \mathllap{w'} = \cos(\phi')[v_x' \cos(\theta') + v_y' \sin(\theta')] \\ + v_z' \sin(\phi')
  \end{aligned}
\end{align}

\noindent where $\phi'$ and $\theta'$ are the latitudes and longitudes after the phase and inclination rotations, and $\delta$ is the orbital inclination. 

We simulate emission spectra for Upsilon Andromedae~b at 12 orbital phases from 0 to 1 (see Figure \ref{fig:drawing}) for the low gravity clear case and the low gravity cloudy case, at the measured inclination of 24$^\circ$ \citep{Piskorz_2017}. Initial calculations showed that the low gravity clear case and the high gravity clear case had similar atmospheric structures and spectra, as explained in \S~\ref{sec:Results}, and so we subsequently only re-grid and simulate spectra for the low gravity clear and low gravity cloudy cases. Then, to show the general effects of how inclination changes manifest in high-resolution spectra, we create a grid with 8 phases from 0 to 1, and 5 inclinations from 90$^{\circ}$ to 0$^{\circ}$ for the low gravity clear and cloudy cases.

\subsection{Simulating Emission Spectra}\label{subsec:1}

We post-process the GCM results and simulate high-resolution (R = 100,000) emission spectra from 2.308 to 2.314 ~$\mu$m. At this resolution, the individual lines that make up the CO bandhead at $\sim$ 2.290~$\mu$m can be distinguished. We choose this K band wavelength range to overlap with the coverage of high-resolution instruments such as CRIRES+ and IGRINS \citep{Kaeufl2004,Marsh2007}. The physical behavior we observe here, especially Doppler effects from atmospheric circulation and rotation, should generally also be applicable at other wavelengths as well.

From the GCM output, we have temperature, pressure, atmospheric optical depth across the grid cell, and cloud properties at each grid cell. Based on line-of-sight vectors to the observer, we divide the atmosphere into columns, and then apply a radiative transfer model to calculate the emergent light intensity at each grid cell on the observer-facing hemisphere. We follow the procedure from \cite{2017ApJ...851...84Z} for the geometry of post-processing the GCM results, but with a significant upgrade of the radiative transfer scheme.

In order to model Doppler shifts from circulating winds and planet rotation we calculate the line of sight velocity as

\begin{align}
\phantom{i + j + k}
&\begin{aligned}
\mathllap{\rm{v_{LOS}}} &= u' \rm{sin}(\theta') + \\
&\qquad v' \rm{cos}(\theta')\rm{sin}(\phi') - \\
&\qquad w' \rm{cos}(\theta') \rm{cos}(\phi') + \\
&\qquad \rm{cos}(\pi/2-\delta) (\Omega)(R_p+z) \rm{sin}(\theta')\rm{cos}(\phi') 
\end{aligned}
\end{align}

\noindent where $\Omega$ is the planetary rotation rate. Throughout this paper, we neglect the Doppler shifts from orbital motion in the emission spectra, as these will be constant for the entire planet surface and can be subtracted from observed emission spectra.

To create a final set of spectra we follow the procedure from \cite{2017ApJ...851...84Z}. For each planet model we calculate four different spectra: without Doppler shifting, with only the wind contribution to Doppler shifts, only the rotational contribution, and with Doppler shifts from the combined influence of winds and rotation. To characterize the aggregate effect of winds and rotation on the emission spectra, we cross correlate each spectrum against a base case without Doppler effects and fit the cross correlation to a Gaussian. This allows us to evaluate the respective contributions of winds and rotation, as well as their combined effect. This window into the winds of the planet comes from our ability to analyze the Doppler shifts of the spectra and disentangle the wind and rotational contributions to line-of-sight motions. Similarly to \cite{harada2021}, we take the net Doppler shift for each phase and inclination to be the velocity shift of the peak of the cross correlation function.

\subsection{Radiative Transfer Upgrades}
In order to include the effects of inhomogeneous aerosols and scattering within the atmosphere, we implement the two-stream radiative transfer scheme of \cite{1989JGR....9416287T}, which we can now use for both the clear and cloudy models. From the analytical matrix solution, we calculate the emergent intensity from each line-of-sight column in the atmosphere, and treat each wavelength bin independently. The two-stream solution reduces to the theoretically exact result in the absorbing only limit, and has accuracy within 10$\%$ when scattering effects play a significant role in atmospheric energy transfer \citep{1989JGR....9416287T}. We show test cases with our upgraded routine under idealized and fully 3-D conditions in the Appendix, demonstrating the correct implementation of these substantial changes.

Each 1-D atmospheric column is divided into discrete layers, parameterized at each layer by temperature, total optical depth of the layer ($\tau_{tot}$), the single scattering albedo ($\varpi_0$), and the asymmetry parameter ($g_0$). Each cloud species contributes an individual aerosol optical depth ($\tau_{cloud, i}$). The total optical depth of each grid cell is a sum of the individual cloud species, as well as the gas optical depth contribution. Within our models, clouds contribute to the single scattering albedo, the asymmetry parameter, and the total optical depth in each layer.

As in the previous works, we assume that the four species have characteristic particle radii of 0.2~$\mu$m, with a log normal distribution and a variance of 0.1 ~$\mu$m and use Mie theory to compute the scattering properties \citep{1984A&A...131..237D, MISHCHENKO1999409}. In order to scale from the 5.0~$\mu$m values calculated for the IR channel of the double gray GCM model to the 2.3~$\mu$m values needed for the wavelength range we are calculating here, we follow the same approach as \cite{2019ApJ...872....1R} and \cite{harada2021}. We assume spherical particles and calculate the cloud opacities as 

\begin{equation}
    \kappa_\text{cloud, i} = \frac{\tau_{cloud, i}}{ds} \frac{Q_{2.3\mu\text{m}}}{Q_{5.0\mu\text{m}}}
\end{equation}

\noindent where $\frac{Q_{2.3\mu\text{m}}}{Q_{5.0\mu\text{m}}}$ is the ratio of the extinction efficiencies at 2.3 ~$\mu$m and 5.0 ~$\mu$m, and $\kappa_{cloud}$ is given here in units of cm$^{-1}$. 

Following \cite{harada2021}, we calculate the line-of-sight total layer optical depth as 

\begin{equation}
\tau_{\lambda} = \int(\kappa_{gas} + \kappa_{cloud, \ total}) \ dl
\end{equation}

\noindent where $\kappa_{gas}$ is the gas opacity and $\kappa_{cloud}$ is the cloud opacity component. The optical depth is calculated for each wavelength bin, as the $\kappa_{gas}$ is a wavelength-dependant function. Following \cite{2017ApJ...851...84Z}, the gaseous opacities are calculated assuming thermochemical equilibrium for a solar composition mixture and we assume that cloud opacities are wavelength independent over the narrow 2.308-2.314~$\mu$m wavelength range considered in this work as in \cite{harada2021}.

Next, we calculate the single scattering albedo and asymmetry parameters, $\varpi_0$ and $g_0$ respectively, for each layer. The single scattering albedo ranges from 1 for fully conservative scattering to 0 for full absorption. The asymmetry parameter defines the direction of aerosol scattering, with 1 being complete forward scattering, 0 being isotropic scattering, and -1 being complete backward scattering. We assume that $\varpi_0$, and $g_0$ do not vary with wavelength over the narrow wavelength range we use. We calculate $\varpi_0$ and $g_0$ from their weighted sum in each layer:

\begin{equation} \label{eqn_w0}
\varpi_0 = \sum \frac{\tau_{cloud, i}}{\tau_{tot}} \times \varpi_{0,i}
\end{equation}

\begin{equation} \label{eqn_g0}
g_0 = \sum \frac{\tau_{cloud,i}}{\tau_{tot}} \times g_{0,i} .
\end{equation}

The treatment of the radiation field is split into two components: one from the planet's thermal emission and one from the scattering of incident starlight. At each wavelength, we sum the two components and compute the upwards and downwards intensity, where upwards is defined in the direction toward the observer, not  the radial distance from the planet. \cite{1989JGR....9416287T} refers to these two components as the optical and infrared contributions, and confines each treatment to its respective wavelength band. This approach works for cooler planets; however, for hot Jupiters like Upsilon Andromedae b, the scattered light and thermally emitted components contribute at overlapping wavelengths. For this reason, we track both components at each wavelength of our radiative transfer calculation and sum them to determine the total (scattered plus thermally emitted) emergent flux. We calculate the optical contribution using the quadrature scheme and the infrared contribution using the hemispheric mean scheme from \cite{1989JGR....9416287T}\footnote{Of note, the original two-stream solution from \cite{1989JGR....9416287T} has a typographical error in Equation 42 for the source function technique. The corrected form implemented here is: $E_l = [C_{n+1}^+(0) - C_{n}^+(\tau_n)]e_{2n+1} + [C_{n}^-(\tau_n) - C_{n+1}^-(0)]e_{4n+1}$}.

To compute the total flux seen by an observer, we sum together the emergent fluxes coming from each latitude-longitude grid cell on the observer-facing hemisphere weighted by the grid cell's sky-projected area. Implicit to our two-stream implementation is that each 1-D atmospheric column is independent of the others i.e.\ there is no net loss or gain of scattered photons between adjacent columns. This should be a reasonable assumption, especially considering that the two-stream approach is already an approximate solution to the full scattering problem. A more physically accurate Monte Carlo radiative transfer calculation would be extremely computationally expensive and is beyond the scope of our current work \citep{lee+2017, 2019MNRAS.487.2082L}.

In addition to being able to include the effects of aerosol scattering and absorption, a second significant improvement of this two-stream radiative transfer approach over our previous work is that we can now accurately incorporate the reflected component of incident starlight into the planetary spectra. Especially in cloudy models, reflected starlight can have a significant contribution to the total flux seen by an observer, even potentially at the IR wavelengths modeled in our current work. To model the incident starlight (a required input to the scattered light component of the two-stream calculation), we calculate the intensity normal to the planet's surface at each grid cell as

\begin{equation} \label{eqn_normal_intensity}
I \cdot \hat{\VF{n}} = \sin(\phi) \sin(\zeta) + \cos(\phi) \cos(\zeta) \cos(\theta)
\end{equation}

\noindent where $\zeta$ is the system obliquity. Additionally, we assume that the spectrum of the starlight is a Planck function with a characteristic temperature equal to that of the host star. Further work could improve on our results by parameterizing the host star spectrum more accurately, but this is expected to have a minor effect over the limited wavelength range we are modeling here and is beyond the scope of this paper.

To ensure a degree of self-consistency, this two-stream radiative transfer technique is the same as the one used within the GCM, as elaborated in \S~\ref{subsec:3d}, albeit we apply a double-gray approximation in the GCM and calculate radiative fluxes in just two bands: one for the IR wavelengths and one for the visible wavelengths. The two-stream approximation is accurate for our uses in the GCM, in which the main concern is computing the flux divergences and consequent heating rates. However, when calculating spectra we must solve the radiative transfer equations at each wavelength.

\section{Results}\label{sec:Results}
As far as we are aware, our analysis of Upsilon Andromedae b is the first time simulated high-resolution spectra have been calculated from 3D atmospheric dynamics models for non-edge-on orbits. First, we demonstrate our methods by simulating the atmospheric structure and high-resolution emission spectra for the specific case of Upsilon Andromedae b. Next, we use the GCM results for Upsilon Andromedae b as a base case and simulate how changes in inclination and phase manifest in the high-resolution emission spectra more generally. Throughout this work we compare models with low or high gravity and with and without clouds.

\subsection{GCM Simulations of Upsilon Andromedae b}\label{subsec:structure}

\begin{figure*}[!htb]
\begin{center}
\includegraphics[width=0.9\linewidth]{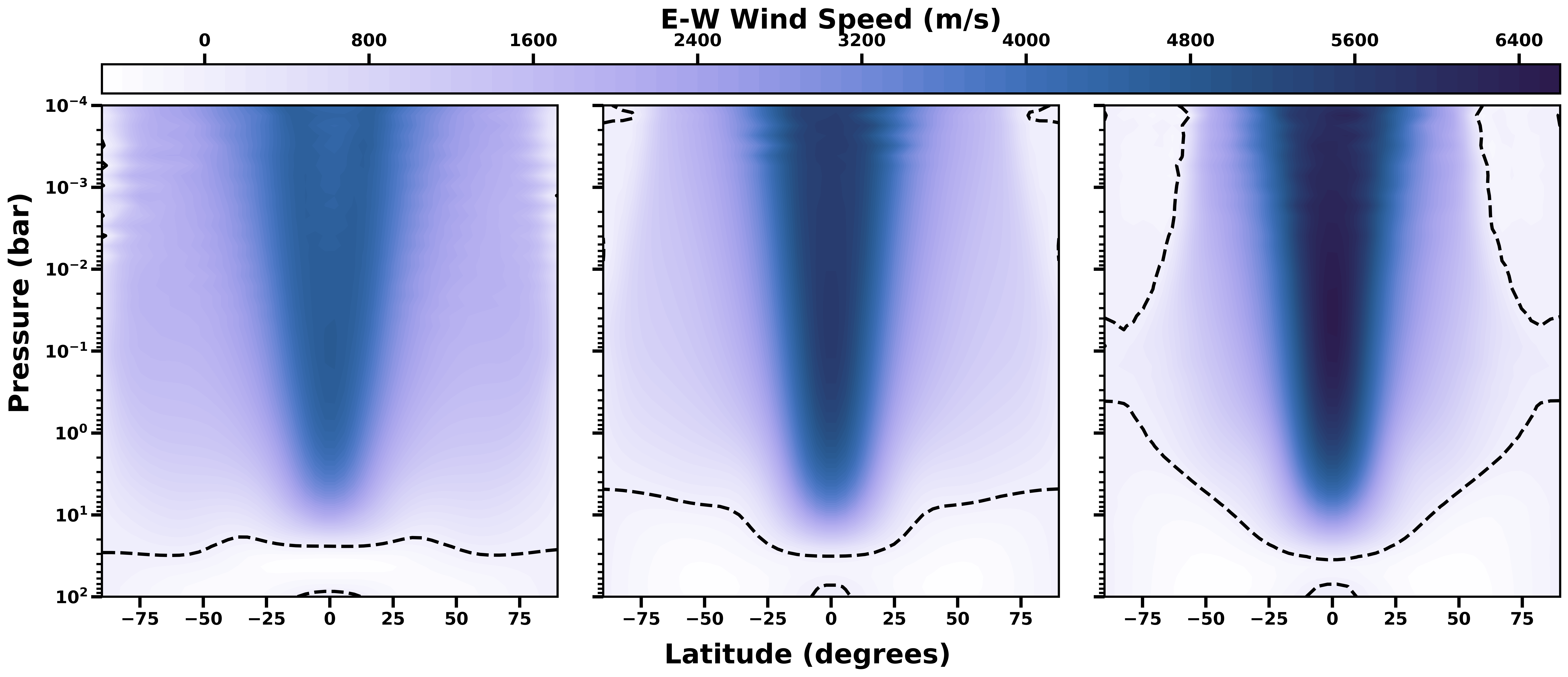}
\caption{Longitudinally averaged zonal wind speeds of Upsilon Andromedae b throughout the atmosphere. The dotted black contour lines correspond to average wind speeds of 0 km/s. The dark central band corresponds to the equatorial jet, which extends to pressure levels of approximately 10 bar. From left to right the panels correspond to the high gravity clear case, the low gravity clear case, and the low gravity cloudy case.}
\label{fig:winds}
\end{center}
\end{figure*}

\begin{figure*}[!htb]
\begin{center}
\includegraphics[width=0.9\linewidth]{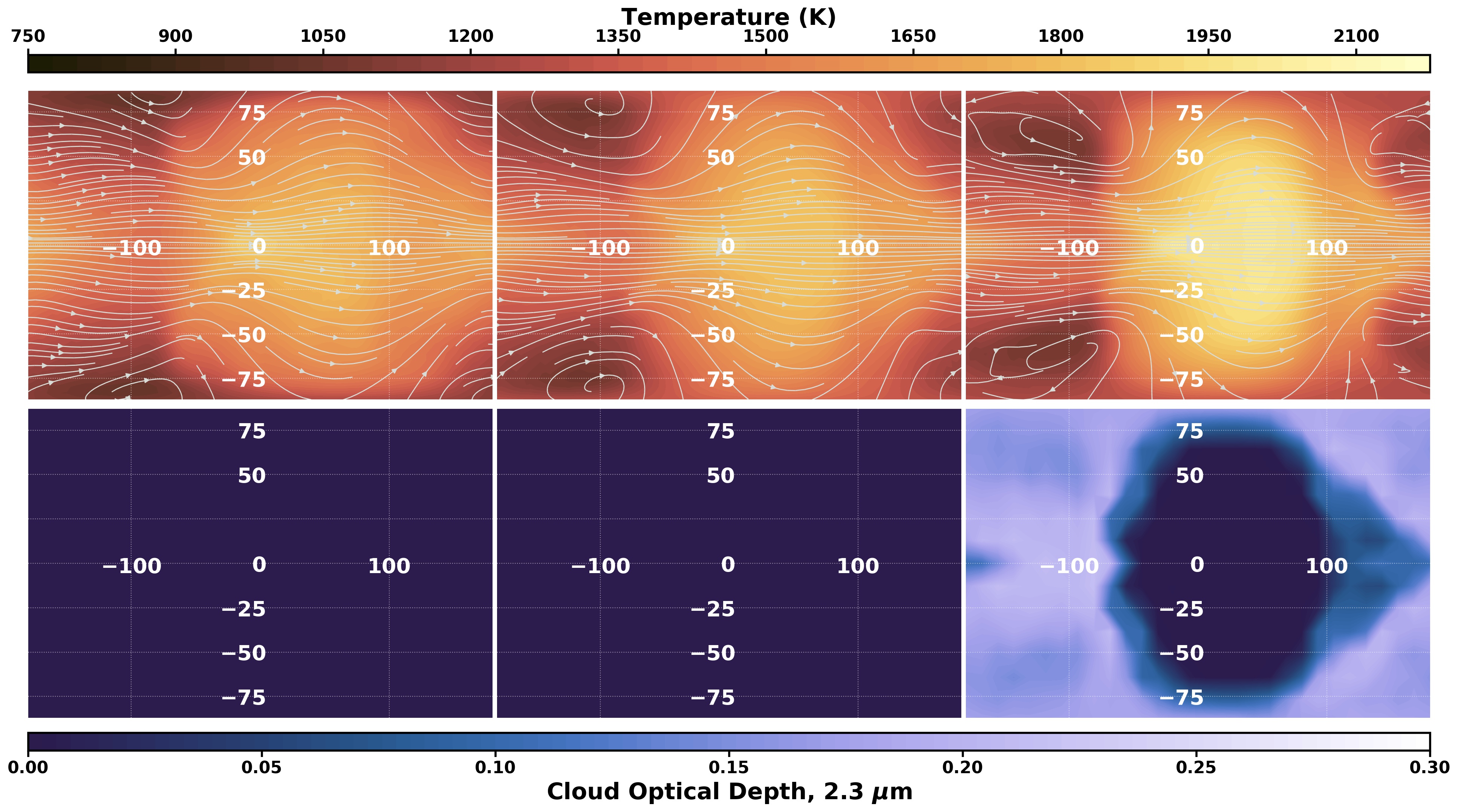}
\caption{The temperature, wind, and cloud structure of Upsilon Andromedae b at a pressure level corresponding to approximately the IR photosphere (90 mbar for the low gravity models and 153 mbar for the high gravity model). The top panels show a cylindrical projection of the temperature and wind for the three GCMs. The magnitude and directions of the winds are shown with the vector contours, with denser line distributions corresponding to faster wind speeds. From left to right the panels correspond to the high gravity clear case, the low gravity clear case, and the low gravity cloudy case. The bottom panels show the integrated cloud optical depth from the given pressure level to the top of the atmosphere.} 
\label{fig:phot}
\end{center}
\end{figure*}

\begin{figure*}[!htb]
\begin{center}
\includegraphics[width=0.9\linewidth]{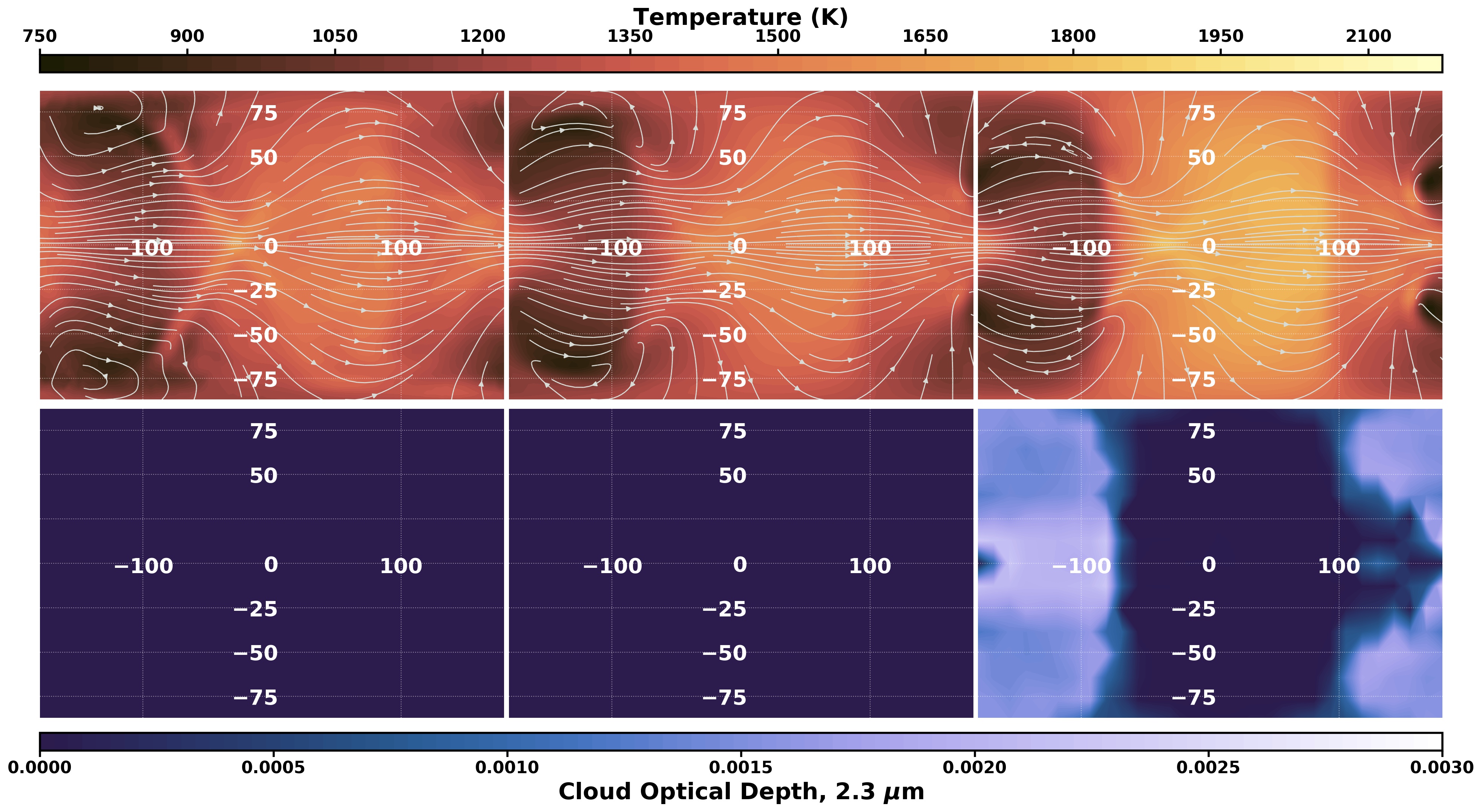}
\caption{The temperature, wind, and cloud structure of Upsilon Andromedae b at a pressure level of 1 mbar. As in Figure \ref{fig:phot}, the panels from left to right correspond to the high gravity clear case, the low gravity clear case, and the low gravity cloudy case. Of note, the scaling of the cloud optical depth maps here are not identical to Figure \ref{fig:phot} and the integrated cloud optical depth is from 1 mbar to the top of the atmosphere here.} 
\label{fig:1mbar}
\end{center}
\end{figure*}

\begin{figure*}[!htb]
\begin{center}
\includegraphics[width=0.9\textwidth]{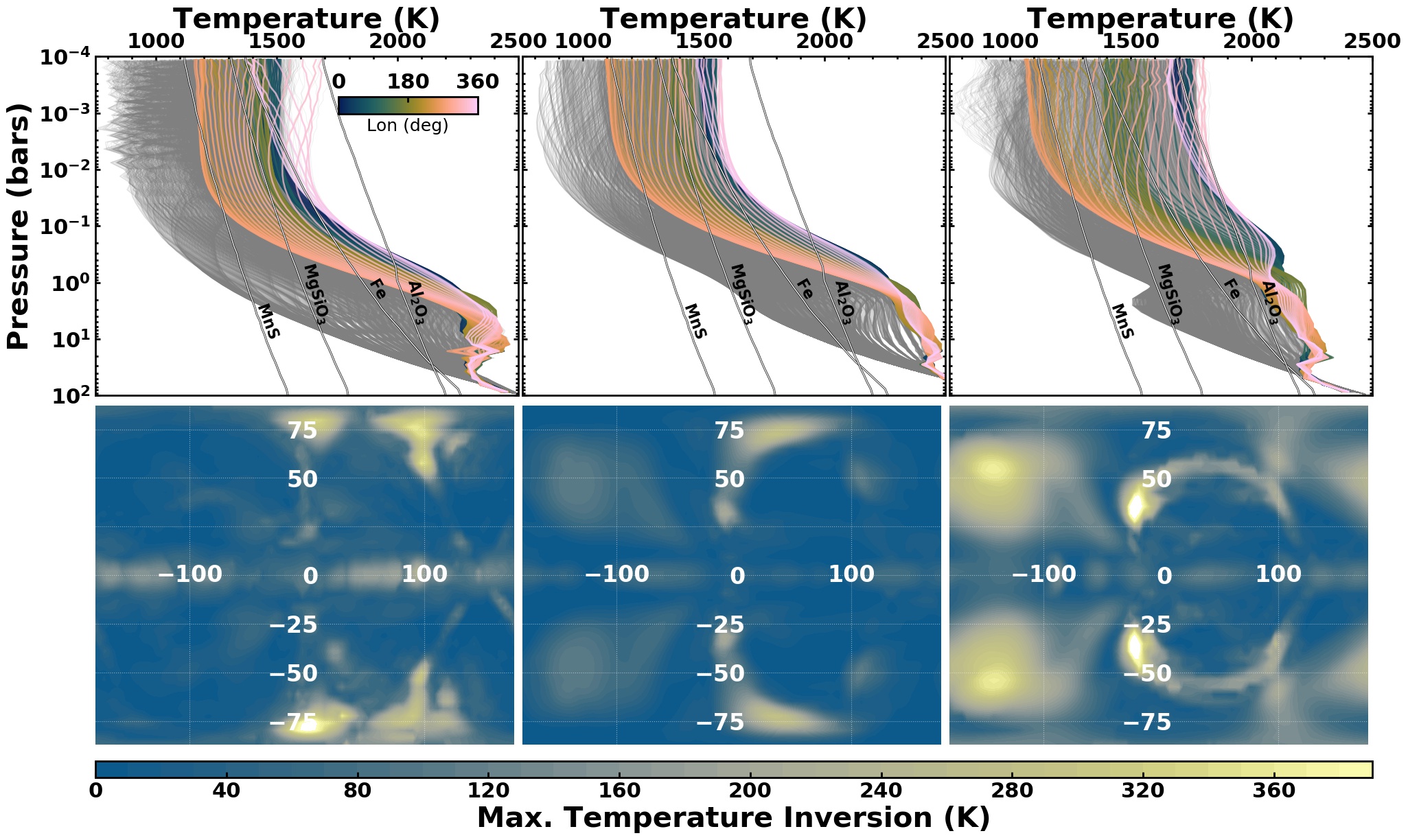}
\caption{Pressure-temperature profiles for Upsilon Andromedae b for our three different models. The colored lines correspond to the equator (with the longitude values indicated by the colorbar), while the gray lines are the pressure temperature profiles for all other latitudes. The four white sloped lines correspond to the condensation curves for MnS, Al$_2$O$_3$, Fe, and MgSiO$_3$. From left to right the panels correspond to the high gravity clear case, the low gravity clear case, and the low gravity cloudy case. The bottom three panels are the maximum temperature inversions at each latitude and longitude (the vertical and horizontal axes, respectively). In the cloudy case there is a tendency for the regions of strongest temperature inversion to occur in regions slightly west of the substellar point (slightly $<$~360 degrees longitude), the hottest profiles are generally tens of degrees eastward of the substellar point ($\sim 40^\circ$, see also figures \ref{fig:phot} and \ref{fig:1mbar}), due to the eastward circulation.}
\label{fig:tp}
\end{center}
\end{figure*}

Figures \ref{fig:winds} $-$ \ref{fig:tp} show the output of the GCM models in terms of the temperature, wind, and aerosol structure. The atmospheric structure of our GCM results have the same standard hot Jupiter properties as detailed in previous work \citep[e.g.][]{2002A&A...385..166S, 2019ApJ...872....1R}. In short, the clear models show a circulation pattern dominated by an eastward equatorial jet, which advects the hottest region of the atmosphere east of the substellar point. The cloudy model maintains this equatorial jet and associated eastward advection of gas heated on the dayside. However, the temperature profile of the atmosphere contains more complex substructure as it is strongly influenced by feedback with the formation and dissipation of cloud species at their condensation points. The different chosen values for gravity within the GCM alters the scale height, which allow the solar radiation to penetrate to greater pressures in the high gravity cases. However, the total integrated heating remains essentially the same. Additionally, the higher gravity also results in a smaller Rossby deformation radius which has a modest effect on the circulation dynamics.

Figure \ref{fig:winds} shows that all clear and cloudy models of Upsilon Andromedae b have an eastward flow dominated by an an equatorial jet. The equatorial jet can have wind speeds of around 6 km/s, extending approximately 20$^{\circ}$ North and South. Near the poles, the winds are much slower, but still generally follow the eastward flow present in the rest of the atmosphere and usually have speeds less than 1000 m/s, significantly less than the equatorial winds. Furthermore, it is worth noting that these regions will be more visible for non-transiting geometries. Slightly noticeable in the upper atmosphere are vertical variations in the wind speeds; a similar pattern appears in the temperature structure as well (Figure \ref{fig:tp}). These are numerical effects, due to some combination of artificial wave reflection off the top boundary and under-resolving the physics in this region with very short radiative and dynamical timescales. A sponge layer in the uppermost layers (Beltz et al., in prep.) reduces this noise, but careful tests \citep{harada2021} show that they do not influence our simulated spectra.

Clouds scatter and reflect light and change the average location from which observed photons originate; the inhomogeneous cloud coverage in these models also makes the photospheric pressure highly location-dependent. Figure \ref{fig:phot} shows the temperature and wind structure of Upsilon Andromedae b at a horizontal layer in the upper atmosphere (80 mbar for the low gravity models and 153 mbar for the high gravity model). At 80 mbar the cumulative optical depth from clouds is between 0 and 0.25, meaning that the clouds are not yet optically thick but still have a significant impact on the radiative transfer in the upper atmosphere. For the clear models, these pressures correspond to the IR photosphere for the double-gray GCM: the location from which the photons that make up the continuum flux emerge. However, when clouds are present and optically thick, there is no single pressure level that corresponds to the photosphere \citep{roman2021clouds}. Analyzing the pressure level where the continuum forms becomes far more complex when clouds are present \citep{harada2021}.

The detailed strength and shapes of the spectral lines in disk-integrated emission spectra will be determined by the cloud and temperature structure across these pressure ranges. Figure \ref{fig:1mbar} shows the temperatures and winds at 1 mbar. At this pressure level, we are above most of the clouds present in the cloudy model, although some lingering, optically thin clouds remain on the nightside at high latitudes. In most locations, the temperature at 1 mbar is significantly colder than at the IR photosphere. Absorption lines will be present in the high-resolution spectra from locations where the upper atmosphere is colder than the continuum flux pressure level, while emission lines will be present when the upper atmosphere is hotter than the continuum flux pressure level. This is because the pressure corresponding to where line cores form is lower than the pressure corresponding to the continuum flux.

We find that there are stronger temperature inversions in the upper atmosphere of Upsilon Andromedae b when clouds are present. We quantify a thermal inversion similarly to \cite{harada2021}: the maximum continuous increase in temperature across contiguous vertical levels at a single latitude and longitude location. The thermal inversions are largest for the cloudy models, although we find small temperature inversions in the clear GCMs as well simply due to atmospheric dynamics. The clouds scatter and absorb starlight near the top of the cloud deck. This raises the temperature of the upper atmosphere clouds and also insulates the lower atmosphere from incident radiation, lowering the temperature. These temperature inversions can be seen in Figure \ref{fig:tp}. The largest temperature gradients and inversions occur at pressure levels from approximately 10$^{-4}$ to 10$^{-1}$ bar, which corresponds to locations above the top of the cloud decks. We elaborate on how this behavior influences the emission spectra in section \ref{subsec:highres}.

The formation and dissipation of clouds on hot Jupiters is dependant upon the atmospheric temperature profiles, compared to the condensation curves of each species. At many points throughout the atmosphere, but predominantly on the cool nightside, temperatures drop below the condensation curves of MnS, Fe, MgSiO$_3$, and Al$_2$O$_3$ and these cloud species form. As the equatorial jet advects cold air from the nightside to the dayside, the temperature rises and clouds dissipate. The presence and effects of temperature inversions in our models are broadly similar to those described in \cite{harada2021}, who were the first to discuss cloud-induced inversions in the GCMs of \cite{2019ApJ...872....1R}.

Figure \ref{fig:tp} shows that temperature inversions are predominantly on the dayside for the clear models, but occur on both the day and the nightside for the cloudy model. The largest temperature inversions for the clear models are near the poles, and have a maximum magnitude of 400 K. However, the clear models also display smaller temperature inversions, of approximately 200 K, along the equator. In contrast, the cloudy model has both larger temperature inversions (of up to 500 K), as well as temperature inversions covering a larger area of the planet.

\subsection{High-Resolution Emission Spectra of Upsilon Andromedae b}\label{subsec}

Here we present simulated spectra for the low-gravity clear and cloudy GCMs, which allow us to identify how the inclusion of clouds manifests in high-resolution emission spectra of non-transiting planets. \citep[A similar assessment for transiting planets was performed in][]{harada2021}. We also calculated simulated spectra for the high-gravity model, but found only minor differences between that and the low-gravity model. As shown in Figures \ref{fig:winds} - \ref{fig:tp}, there are minimal differences in the \mbox{3-D} structure between these two cases; mostly the higher gravity shifts the depth at which stellar light is deposited and circulation driven, but the photospheric emission is shifted in similar way, such that there is minimal influence on the resulting spectra.

\begin{figure*}[!htb]
\begin{center}
\includegraphics[width=0.9\linewidth]{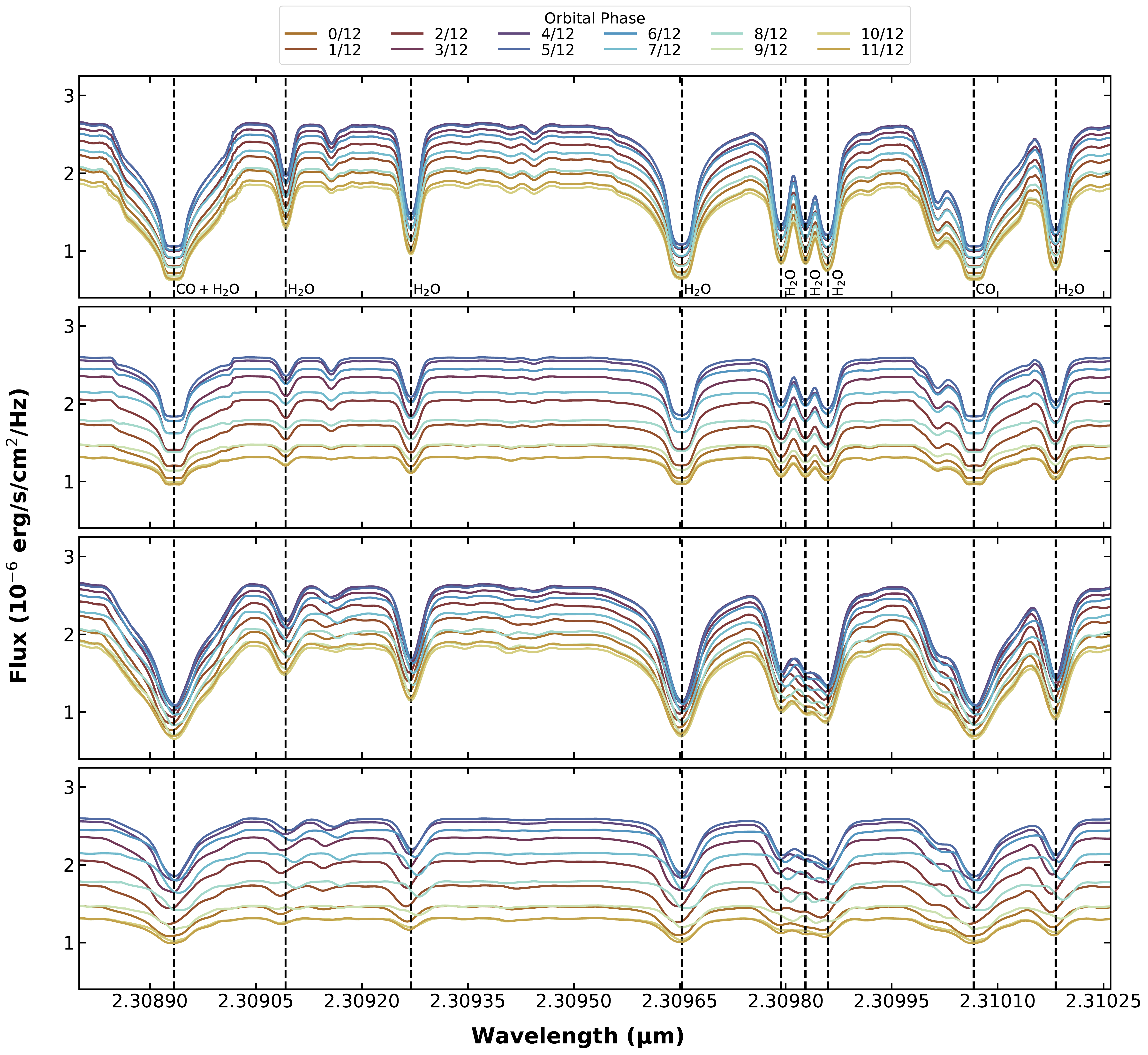}
\caption{Simulated disk integrated emission spectra for Upsilon Andromedae b at 12 equally spaced orbital phases, as viewed at its orbital inclination of 24$^\circ$. The fractional orbital phases correspond to 0.0, 0.083, 0.167, 0.25, 0.333, 0.417, 0.5, 0.583, 0.667, 0.75, 0.833, and 0.917.} From top to bottom, the panels are: models with Doppler effects off (first clear, then cloudy) and models with Doppler effects on (first clear, then cloudy). The vertical dotted lines indicate specific absorption features (as labeled) in the rest frame.
\label{fig:spectral-lines}
\end{center}
\end{figure*}

Figure \ref{fig:spectral-lines} shows the simulated high-resolution spectra of Upsilon Andromedae b for 12 equally spaced orbital phases between 0 and 1, with Doppler effects turned on and off, respectively and as viewed at an orbital inclination of 24$^\circ$. We show a subset of the total simulated wavelength range, from 2.309~$\mu$m to 2.311~$\mu$m, in order to highlight carbon monoxide and water absorption lines. This matches the wavelength range featured in previous studies of high-resolution emission spectra from our group \citep{2017ApJ...851...84Z, harada2021, Beltz}.

The continuum flux amplitude and absorption features reveal the underlying temperature structure of the planet. The shape and depth of the absorption lines are in large part set by the vertical temperature structure, tracing the difference in temperature between the depths at which the line cores become opaque and the deeper regions from which the continuum emission emerges. Each location on the planet may produce a different emitted spectrum, due to the 3-D spatial structure of the atmosphere, which are then combined into a single disk-integrated spectrum, which is what we would observe.

We see absorption features in both the clear case and the cloudy case, although the strength of the lines is significantly larger in the clear case because cloud opacity limits the pressure (and therefore the temperature) range from which different wavelengths emerge. Additionally, many of our models tend to produce flat-bottomed line cores, as the line cores are optically thick in near-isothermal portions of the atmosphere. When we turn off the Doppler shifts in the cloudy model (second panel of Figure \ref{fig:spectral-lines}), there is a hint of emission reversals at the center of line cores, at a few of the orbital phases. This would indicate some weak contribution from regions with temperature inversions to the disk-integrated spectrum.

Doppler effects, from the planet rotational speeds and wind speeds on the order of kilometers per second, change the shape of the simulated emission spectra at this high-resolution. The addition of Doppler shifts preserves the continuum flux amplitude, but introduces important differences from the emission spectra without Doppler effects, broadening the sharpest features and decreasing absorption line depths. Additionally, the lines can have a net red- or blue-shift, depending on how the line-of-sight velocities from planet winds and rotation spatially align with the hotter and brighter regions of the atmosphere, as well as the regions with large thermal inversions. In general the net Doppler shifts in the cloudy model appear to be larger than those in the clear model, while the amount of broadening seems similar between the two. This result is in line with previous work measuring the net Doppler shifts of clear and cloudy hot Jupiters \citep{harada2021}.

The largest difference between the clear and cloudy models is that the addition of clouds decreases the continuum flux level of Upsilon Andromedae b, especially on the planet's nightside. The increase in planetary albedo from the presence of clouds leads to a net reduction in global temperature. Furthermore, the fractional variation between the minimum and maximum continuum fluxes is larger for the cloudy case than for the clear case. The maximum continuum flux is $\sim$40\% above the minimum flux for the clear case, while in the cloudy model the maximum flux is  $\sim$100\% above the minimum. This is due to the fact that the cloudy model has significantly larger upper atmosphere day-night temperature differences than the clear model. These effects are further compounded by the fact that the photospheric pressure is at lower pressure on the nightside relative to the dayside due to clouds forming on the former but not the latter. This behavior is shown in Figure \ref{fig:1mbar}, and consistent with previous results \citep{2019ApJ...872....1R, parmentier2021cloudy, roman2021clouds}. While the amplitude of change is different between the clear and cloudy models, in both cases the peak flux occurs before a phase of 1/2, consistent with the similarity in circulation patterns between the two, with the standard eastward advection of the hot gas away from the substellar point.

\begin{figure*}[!htb]
\begin{center}
\includegraphics[width=0.8\linewidth]{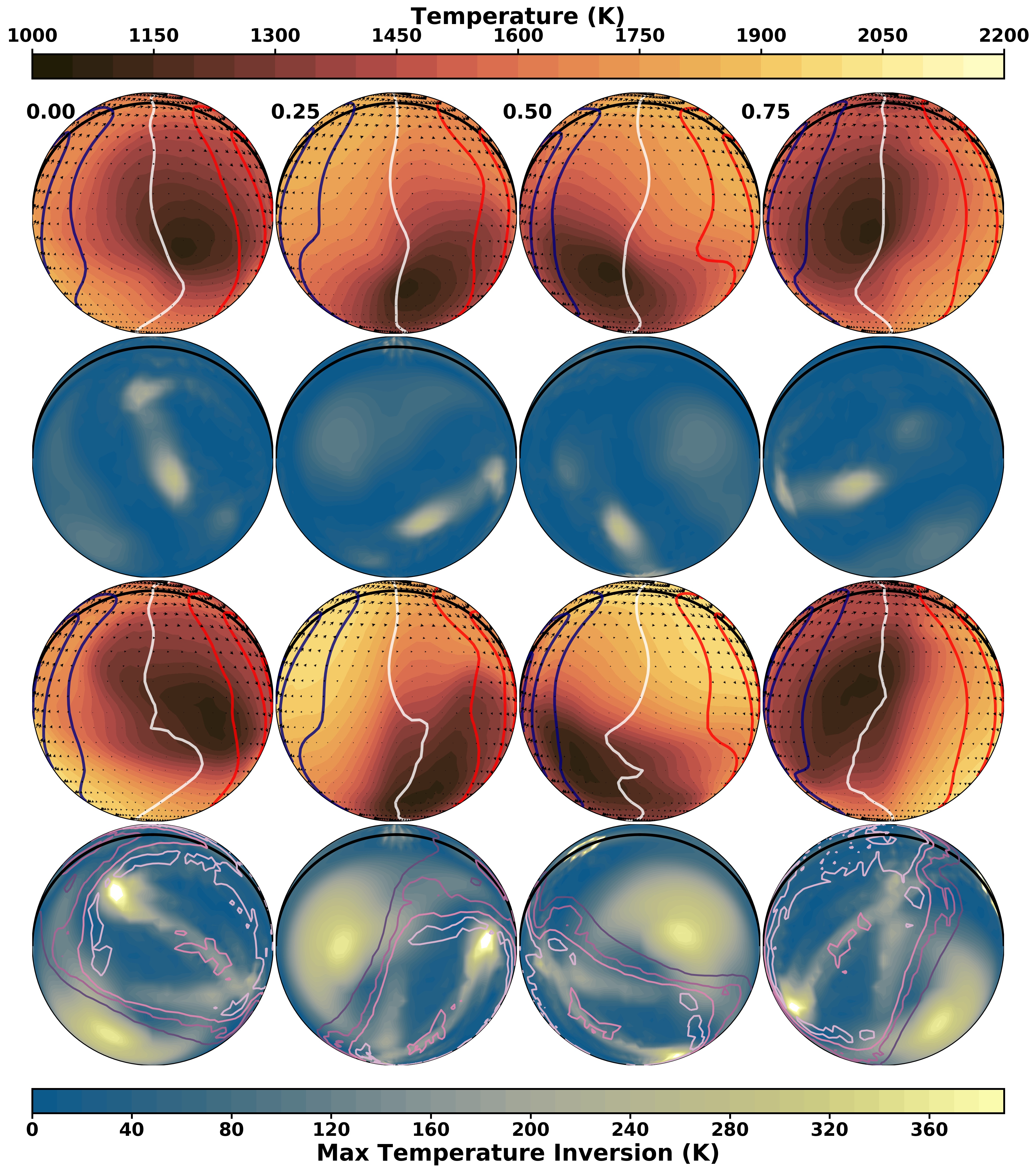}
\caption{Simulated atmospheric structure of Upsilon Andromedae b, viewed at the 24$^{\circ}$ orbital inclination of this planet. The solid black line marks the location of the equator, which is near the edge of the planet disk for this highly inclined orbit. From left to right are maps of the planet at successive phases throughout its orbit (at the phases labeled). The top two rows show the temperature, wind, and line-of-sight velocities at 80 mbar and maps of temperature inversions from the low-gravity clear model, while the bottom two rows show the same set of plots for the low-gravity cloudy model, with contours of cumulative cloud optical depth above 80 mbar added to the last row. From dark to light purple these levels are optical depths of: 0.05, 0.10, 0.15, and 0.20. The line-of-sight velocities take into account both wind and rotational motion; the red and blue contour lines are at $\pm$1, 2, and 3 km/s, while the white line shows zero radial velocity. Depending on the orbital phase and whether the atmosphere is clear or cloudy, we may expect to see emission spectra where the brighter region of the atmosphere contributes a net redshift or blueshift, with this effect perhaps enhanced when clouds cover the cooler regions, and possibly complicated by emission lines from clear regions with temperature inversions.}
\label{fig:hemis}
\end{center}
\end{figure*}

The emission spectra in Figure \ref{fig:spectral-lines} can also be understood in comparison with Figure \ref{fig:hemis}, which shows orthographic projections of the hemisphere of the planet that would be facing the observer, for this planet's inclination and at several orbital phases. In the latter Figure we show temperature and cloud structures near the 80 mbar pressure level, as well as line-of-sight velocities from winds and rotation. While the disk-integrated spectrum will contain information beyond just this slice of the atmosphere, it is instructive for understanding features we observe. For example, while the extreme inclination of this orbit makes most of the southern\footnote{There is an inherent degeneracy between north and south for these systems in general, so it could just as well be the northern hemisphere in view. This would simply appear as a mirror-imaged version of what we see here and would result in effectively identical observed spectra.} hemisphere visible at all phases, the hotter and brighter regions of the clear and cloudy models are more in view at phases from 0.25 to 0.5, corresponding to when we see the largest continuum fluxes in the spectra.

Similarly, Figure \ref{fig:hemis} can help us understand the phase-dependent net Doppler shifts in Figure \ref{fig:spectral-lines}. Focusing in on any of the strong absorption lines, the maximum net blue-shift occurs in-between phases of 0 and 0.25, while the maximum net red-shift is found between phases of 0.5 and 0.75. From Figure \ref{fig:hemis} we can see that these times correspond to when the brightest region of the atmosphere is appearing at the maximally blue-shifted side of the planet or about to disappear over the maximally red-shifted side. \citet{2017ApJ...851...84Z} first identified that the net Doppler shift in high-resolution emission spectra will be weighted by the brightness, visibility, and local line-of-sight velocities of each region of hemisphere in view; here we see an example of this same effect for non-equatorial viewing geometry.

We perform a more rigorous evaluation of the influence of Doppler effects on these simulated spectra by using cross correlation functions (CCFs) to effectively combine the Doppler shifts and broadenings from all of the spectral lines. To do this, we cross correlate two versions of a spectrum that are calculated with and without Doppler effects included, but are otherwise identical (i.e., from the same model and at the same viewing orientation). If the peak of the CCF does not occur at a velocity of zero, this quantifies the net Doppler shift contained within the simulated spectra.

\begin{figure*}[!htb]
\begin{center}
\includegraphics[width=\linewidth]{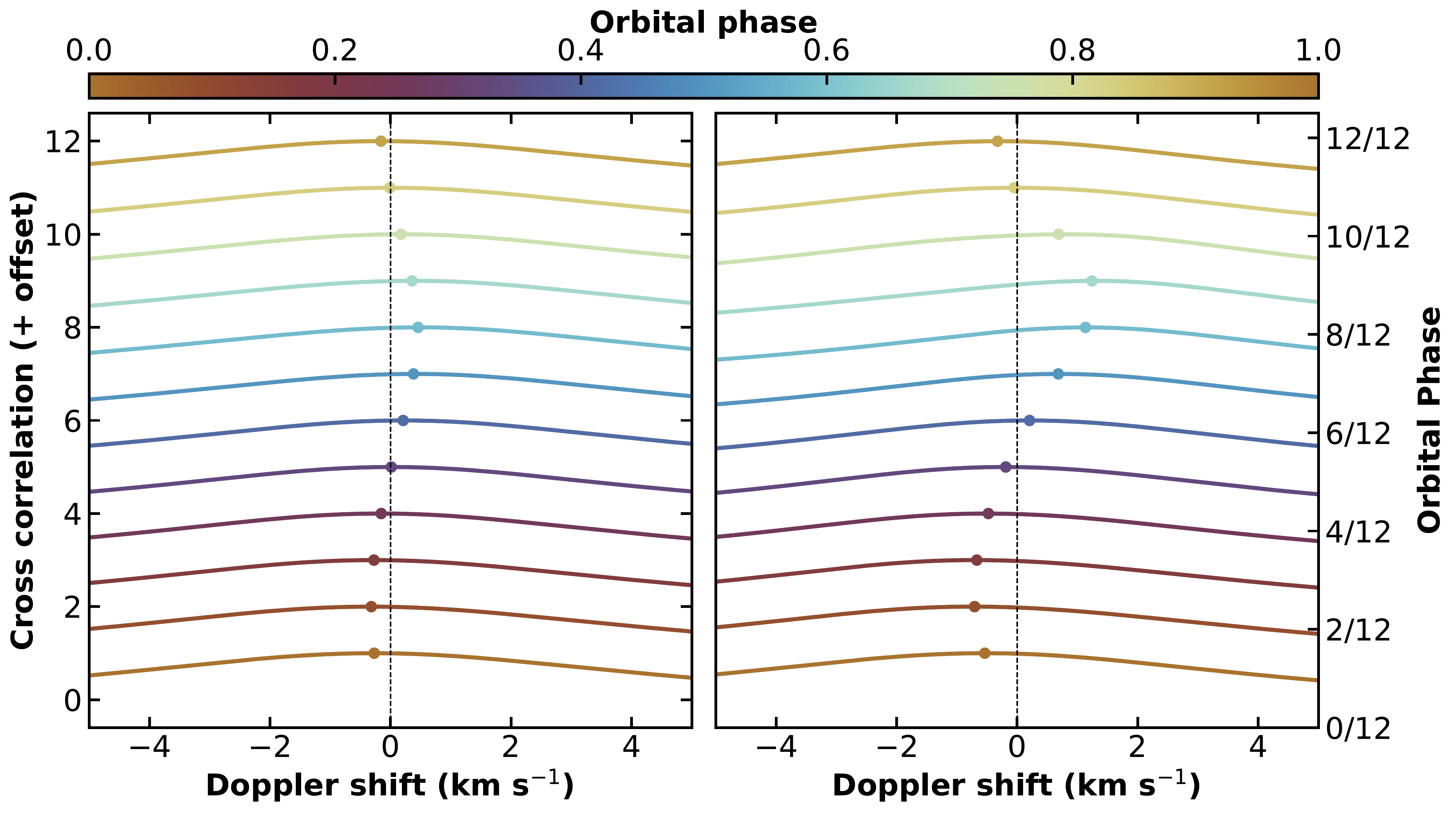}
\caption{We can assess the overall influence of Doppler effects on our simulated spectra by calculating the cross correlation functions (CCFs) between versions of the spectra calculated with and without the Doppler effects from winds and rotation included. Shown here are CCFs of our Upsilon Andromedae b spectra for 12 phases throughout one orbit, normalized and with constant offsets; the low gravity clear case is on the left and the low gravity cloudy case is on the right. The dots mark the peak of each CCF, showing the net Doppler shift imparted to the spectrum at each phase. Overall, the net Doppler shifts are larger for the cloudy models but have a similar trend with phase as the cloud-free models.}
\label{fig:cc}
\end{center}
\end{figure*}

Figure \ref{fig:cc} shows these CCFs as a function of orbital phase for the low gravity clear and cloudy models of Upsilon Andromedae b. Similar to previous high-resolution simulations of clear and cloudy hot Jupiters \citep[][]{2017ApJ...851...84Z, harada2021}, as orbital phase increases from zero, the emission first has a net blue shift and transitions to a net red-shift at a phase of approximately 5/12. The net Doppler shift corresponds to the direction of the winds and planet rotation from the brighter hemisphere. Furthermore, Figure \ref{fig:cc} shows that the net Doppler shifts for the clear models are weaker than for the cloudy models. This is because the presence of clouds leads to large temperature and brightness differences between sections of the visible hemisphere of the planet. The greater temperature heterogeneity means that when the disk-integrated flux is calculated there can be an even more lopsided contribution from the regional Doppler shift of the brighter side compared to the dimmer side, resulting in a greater net red or blue shift.

\subsection{The Effect of Inclination on Hot Jupiter Spectra}\label{subsec:highres}
\begin{figure*}[!htb]
\begin{center}
\includegraphics[width=0.9\linewidth]{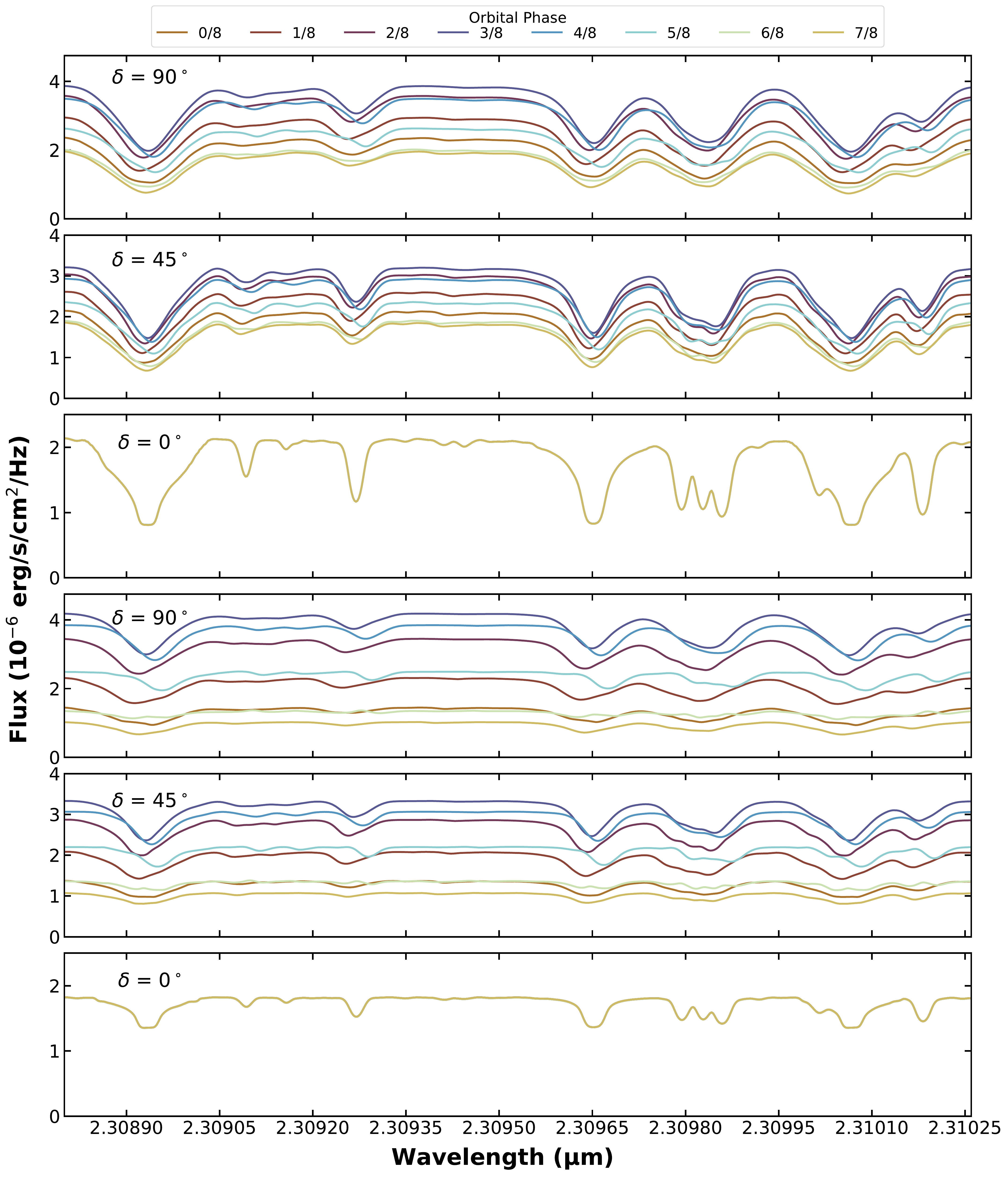}
\caption{Simulated disk integrated emission spectra, with Doppler effects from winds and rotation, for our low gravity clear model (top three panels) and low gravity cloudy model (bottom three panels) at 8 equally spaced orbital phases. For both models we show spectra as viewed at orbital inclinations of $\delta=$ 90$^{\circ}$ (edge-on), $\delta=$ 45$^{\circ}$, and $\delta=$ 0$^{\circ}$ (face-on). We find that the amplitude of all features diminish with decreasing inclination. The fractional orbital phases correspond to 0.0, 0.125, 0.25, 0.375, 0.5, 0.625, 0.75, and 0.875.}
\label{fig:spectral-incs}
\end{center}
\end{figure*}

In order to understand how changes in inclination manifest in high-resolution emission spectra, we remap the low gravity clear and the low gravity cloudy GCMs of Upsilon Andromedae b for 8 phases from 0 to 1 and at 5 inclinations between 90$^{\circ}$ and 0$^{\circ}$. As before, in Figure \ref{fig:spectral-incs} we show a subset of the total simulated wavelength range, from 2.309 ~$\mu$m to 2.311 ~$\mu$m.
For both models, we see three general trends with decreasing inclination: 1) the amplitude of flux differences between spectra at different phases decreases, 2) spectra at all phases become dimmer, and 3) the spectra eventually become identical at an inclination of 0$^{\circ}$. All of these trends can be understood from the changing viewing geometry. As the inclination decreases, we are seeing more polar regions of the planet. This means that there is an increasing area on the planet that remains in view at all phases, and we see less day-night brightness variation. At an inclination of 0$^{\circ}$, the hemisphere of the planet centered on the pole would be visible to an observer throughout the entire orbital phase, and nothing else. Furthermore, the polar regions of the planet are cooler and dimmer than the equatorial regions, decreasing the overall emergent flux.

At these orbital inclinations we see again the same main differences between the clear and cloudy models as we did for the particular viewing geometry of Upsilon Andromedae b. We find that the overall continuum flux is suppressed in the cloudy model (averaged over the orbit), due to the increase in global bond albedo and that the cloudy model has a larger relative variation in continuum flux between phases where the dayside is more in view (around 0.5) and when the nightside is more in view (around 0 or 1), due to the larger global brightness gradients in this model. As before, we also can see noticeably larger Doppler shifts in the spectral lines for the cloudy model.

\begin{figure*}[!htb]
\begin{center}
\includegraphics[width=0.9\linewidth]{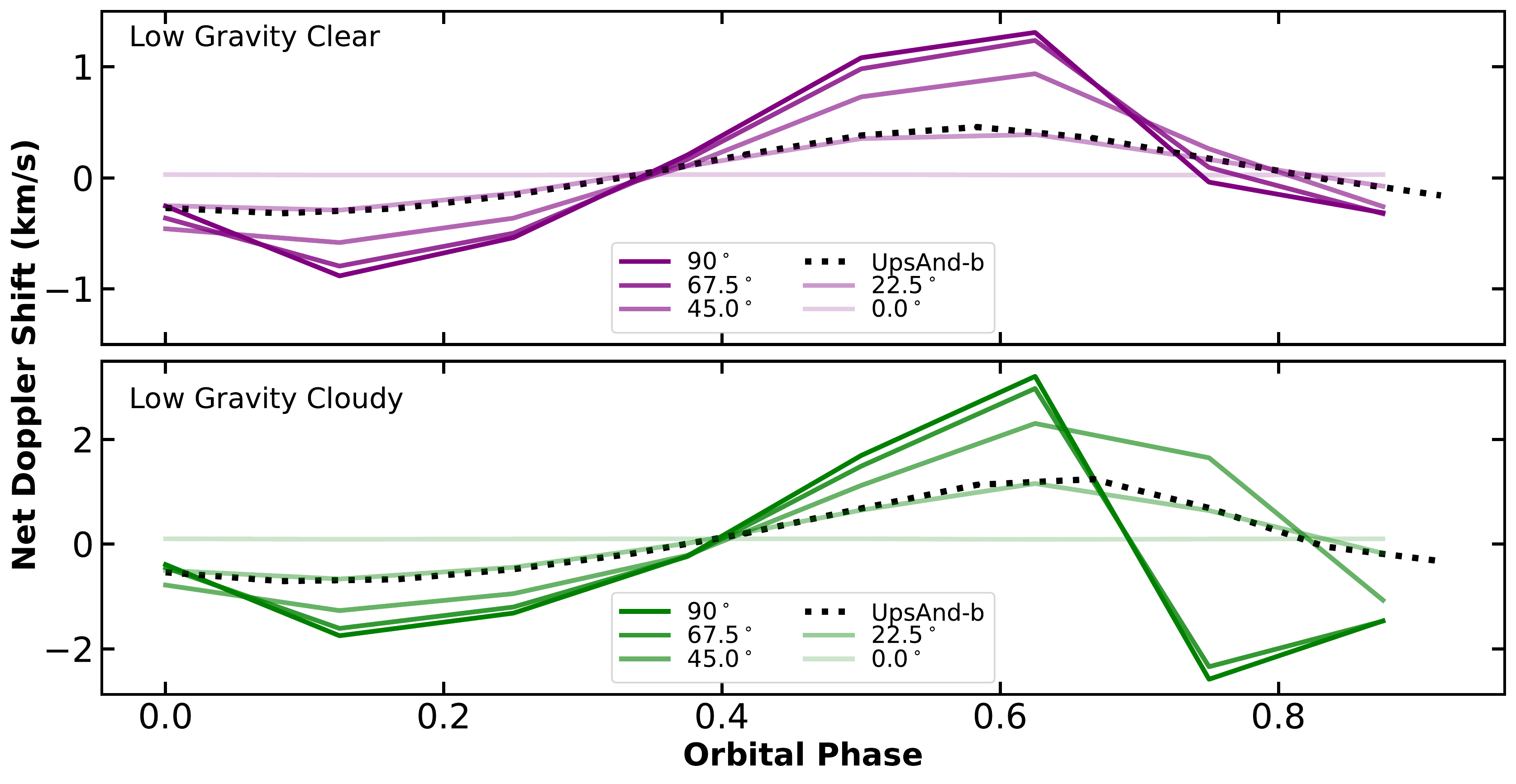}
\caption{The net Doppler shifts in our simulated spectra, calculated from cross correlation functions (as described in the text and with examples shown in Figure \ref{fig:cc}), for our low gravity clear model (top) and low gravity cloudy model (bottom). The dotted black lines are for calculations using the known orbital inclination of Upsilon Andromedae b, while the solid lines show results for inclinations ranging from edge-on (90$^{\circ}$, darkest hue) to face-on (0$^{\circ}$, lightest hue).}
\label{fig:net-dopplers}
\end{center}
\end{figure*}

Figure \ref{fig:net-dopplers} shows the net Doppler shifts as a function of orbital phase for the low gravity clear and cloudy models of Upsilon Andromedae b, as well as clear and cloudy hot Jupiters at inclinations ranging from 90$^\circ$ to 0$^\circ$. Several features are important to note. First, the net Doppler shift decreases with inclination. For both the clear and the cloudy models the net Doppler shifts are at a maximum for edge on inclinations, and approach 0 for 0$^\circ$ inclinations. Second, the cloudy models have net Doppler shifts approximately twice as large as the clear models.

A different circulation pattern, such as the zonal damping discussed in \cite{2013ApJ...762...24S}, could result in different Doppler shift behavior than that discussed in this work. Often the bulk flow in the upper atmosphere will contain a stronger day-to-night flow component, while deeper into the planet atmosphere the circulation pattern will be dominated by the eastward zonal flow \citep[e.g.,][]{RauscherMenou2010}. In this work we focus on emission spectra, which probe deeper into the atmosphere than transmission spectra, and for emission spectra we generally find that we are probing zonal flows \citep[e.g.,][]{2017ApJ...851...84Z} A study of Doppler shifts in high-resolution emission spectra from ultra-hot Jupiters, where magnetic drag may enforce different circulation patterns (Beltz et al., submitted) will appear in Beltz et al. (in prep.).

At an inclination of 90$^\circ$ the maximum net Doppler shift is just over 1 km/s for the clear model, but over 2 km/s for the cloudy model. Again, the source of this difference is due to the larger global brightness gradients in the cloudy model. Last, the presence of clouds complicates the net Doppler shift as a function of orbital phase. At near edge-on inclinations the clear models have CCFs that are quasi-Gaussian and mostly symmetric around a single peak, while the cloudy models can have more complex CCFs, with less symmetric profiles and occasional secondary peaks (as noted in \cite{harada2021}). There is no single source of this behavior, and instead the net Doppler shift is determined by the inhomogeneous cloud distribution, winds, rotation, and pressure-temperature profile.

In addition to the net Doppler shift decreasing with inclination, the width of the CCFs (due to Doppler broadening) also decreases, for both the clear and cloudy models. We fit the CCFs with Gaussian profiles as a rough way to quantify their broadening. Averaged over all orbital phases, the Gaussian widths for the clear model decreased from 5.2 km/s to 3.2 km/s for inclinations from 90$^{\circ}$ to 0$^{\circ}$, while for the cloudy model the widths decreased from 4.8 km/s to 2.9 km/s. However, it should be noted that the CCFs are not always well characterized by a Gaussian, especially for the cloudy models and in particular for orbital phases around 0.75, as the CCFs can show more complex behavior, including multiple peaks. This complexity was also seen in the edge-on cloudy spectra calculated in \citet{harada2021} and can be understood as multiple locations contributing distinct Doppler shifts to the overall disk-integrated spectrum. This occurs near the transition from a net red- to blue-shift as the planet rotates and the blue-shifted component dominates over the red-shifted one. Regardless, for both models and at all phases, as the inclination decreases from edge-on to pole-on, the strength of the Doppler effects decreases, and the CCFs become narrower and centered closer to zero.

\begin{figure*}[!htb]
\begin{center}
\includegraphics[width=0.9\linewidth]{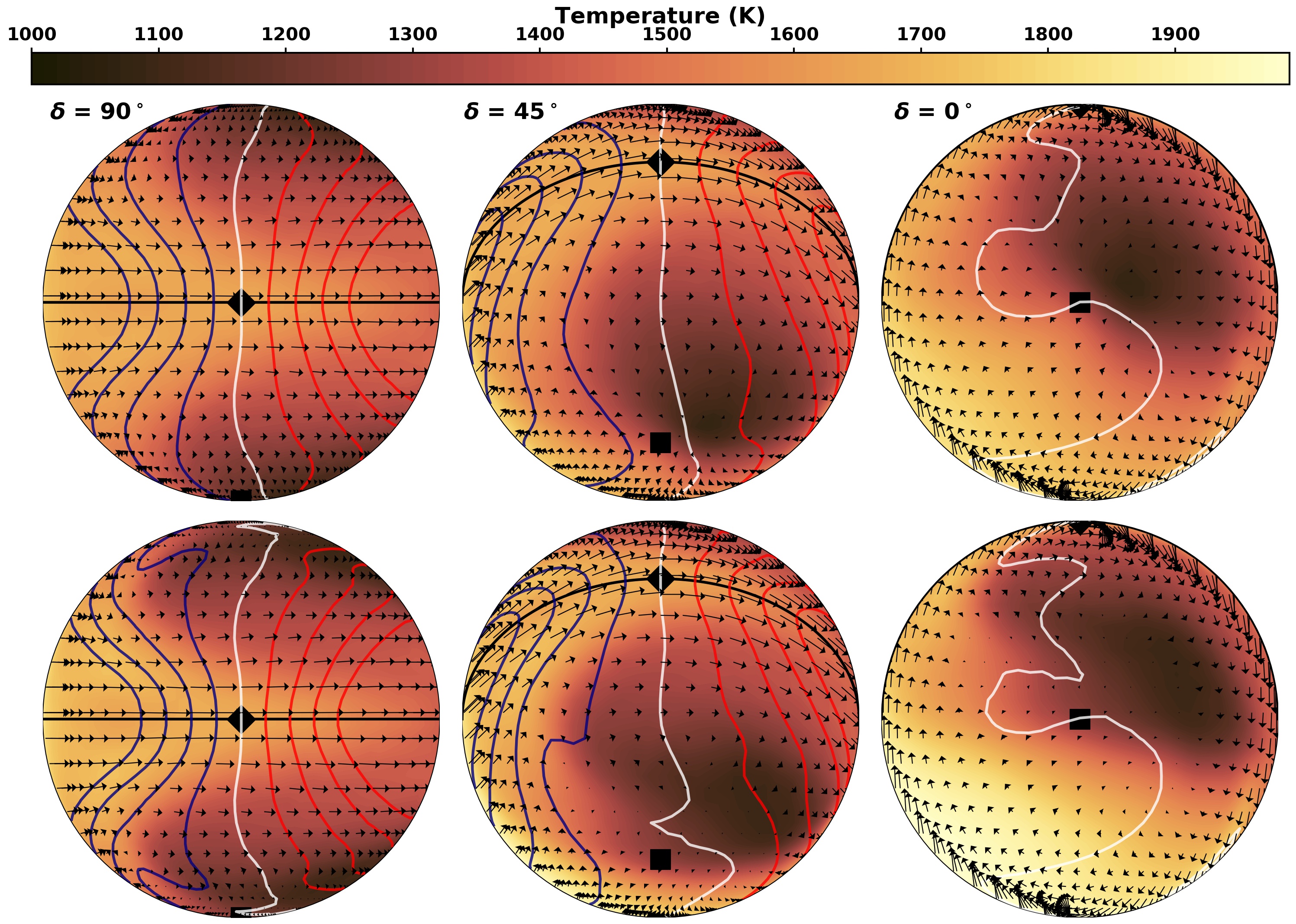}
\caption{Simulated temperatures and wind structures for a hot Jupiter at a pressure level of 80 mbar, for our low gravity clear model (top row) and our low gravity cloudy model (bottom row) at an orbital phase of 0. Also shown are contours of line-of-sight velocities, taking into account both wind and rotation; the red and blue lines are at $\pm$1, 2, and 3 km/s, while the white line shows zero radial velocity. The three panels in each row correspond to orbital inclinations of 90$^{\circ}$ (edge-on), 45$^{\circ}$, and 0$^{\circ}$ (face-on). On each panel the diamond shows the location of the anti-stellar point and the square shows the location of the pole.}
\label{fig:incs}
\end{center}
\end{figure*}

The change in continuum flux amplitude and Doppler effects in the high-resolution emission spectra can also be interpreted in conjunction with Figure \ref{fig:incs}, which shows how the visible hemisphere of the planet changes as inclination decreases from 90$^\circ$ to 0$^\circ$. From an inclination of 90$^{\circ}$ to 0$^{\circ}$, the wind and rotation contributions to Doppler effects decrease for two reasons. First, as the viewing inclination of the planet decreases, the component of planet rotation along the line of sight to the observer decreases. Second, because the dominant planetary winds are in an eastward equatorial jet, the same effect takes place with the Doppler effect of winds. There are still winds around and over the poles, but they are far slower and have a less substantial bulk flow. At 0$^\circ$, the equatorial jet is only visible along the outer edge of the visible hemisphere, and neither it nor the rotation of the planet are contributing to the line of sight velocities. While Figure \ref{fig:incs} only shows the temperature and velocity field at a single level in the atmosphere, the full 3-D calculation of the spectra is strongly influenced by these general trends. One important feature to note is that for all of these simulations the low pressure winds are dominated by the eastward equatorial jet. However, in situations where the zonal jets are damped the winds structure may be dominated by winds flowing from the planet day side to the planet nightside \citep{2013ApJ...762...24S}. These dynamics would in turn alter the Doppler signatures from the bulk air flow, leading to only blue shifts at a phase of 0 and only red shifts at a phase of 0.5.

It is worth noting that the magnitude of the reflected light component of the emergent flux will also decrease as the planet inclination decreases. However, in our model the reflected light component is always negligible compared to the thermal emission component, with further details given in the Appendix. For more highly irradiated and/or higher albedo planets the reflected light component may be a more significant contribution to the total flux at these wavelengths.

\section{Conclusions}\label{sec:conclusions}

The complex three-dimensional atmospheric structure of hot Jupiters can make their measured properties strongly dependent on the orientation with which we view them. This is even more the case when additional sources of inhomogeneity, such as cloud formation in cooler regions, create more drastic differences in physical properties around the planet. Here we present, to the best of our knowledge, the first simulated high-resolution emission spectra calculated for non-transiting viewing geometries. Since atmospheric characterization with high-resolution spectroscopy depends only on the orbital motion of the planet, it works for both transiting and non-transiting systems alike. Our results here consider the role that this non-transiting orientation has in determining the observed planet spectra.

We use the relatively well characterized non-transiting hot Jupiter Upsilon Andromedae b as a representative planet and run GCMs to model its 3-D atmospheric structure. Since the radius (and therefore also gravity) of this planet cannot be constrained by a transit measurement, we run models with higher and lower reasonable values for its gravity. We also simulate models with and without radiatively active clouds. Our models show similar results to previous examples of clear and cloudy atmospheres \citep{2019ApJ...872....1R,harada2021,roman2021clouds}, including: circulation patterns dominated by eastward equatorial jets; cloud formation (if allowed) on the cooler nightside of the planet, causing a larger relative day-night flux difference, and local regional temperature inversions due to the interaction between the cloud structure and atmospheric dynamics. The differences between the high and low gravity models are minor, mostly just a shifting of the atmospheric structure in pressure, but since the levels probed in emission also shift similarly, this has very little influence on the resulting spectra.

In order to accurately include the influence of scattering from clouds in our calculation of high-resolution emission spectra, we have updated our radiative transfer code with a new two-stream line-of-sight radiative transfer treatment. This new routine calculates both the thermal planet emission as well as reflected starlight component and has general applicability well beyond its use in this work. In order to calculate spectra for non-transiting geometries, we also added functionality to our code to tilt the planet at any arbitrary angle, aligning it correctly for the line-of-sight radiative transfer geometry.

Using the Upsilon Andromedae b model as a base case, we show how the physical structure of a planet with a non-edge-on inclination manifests in the emission spectra. We characterize how temperature inversions, winds, and planet rotation create a complex structure of absorption features in the high-resolution spectra. All simulated spectra show net Doppler shifts due to inhomogeneous conditions on the visible disk of the planet and that the amplitude and direction of this shift varies with orbital phase, in agreement with the edge-on models presented in \citet{2017ApJ...851...84Z}.
Overall, we find that the presence of clouds decreases the continuum flux level of the planet and increases the variation of the continuum flux amplitude over the course of the planet's orbit. The cloudy spectra have weaker absorption features as clouds force the emission to emerge from a smaller range of pressures and therefore temperatures.

Furthermore, the spatially dependent pattern of clouds changes the regions of the atmosphere from which we are seeing emission, in complex and inhomogeneous ways. One consequence of this is that the emission spectra from cloudy models have larger net Doppler shifts compared to the clear models, as bright locations around the planet can be more spatially isolated and so their local line-of-sight velocities can contribute more strongly to the disk-integrated spectrum. This behavior was seen in previous simulations of emission for edge-on viewing geometries \citep{harada2021}; here we show that this trend continues even for the highly tilted geometry of Upsilon Andromedae b.

We more generally explore trends in high-resolution emission spectra with changing inclination by using the 3-D structure from our clear and cloudy models of Upsilon Andromedae b as a base case and calculating simulated spectra for a range of inclinations between edge-on and pole-on. We find that for both clear and cloudy models, as the inclination decreases (toward pole-on) the mean continuum flux decreases because the cooler, dimmer polar regions become more visible. The amplitude of variation with orbital phase also decreases, due to the fact that increasingly large sections of the planet remain in the visible hemisphere throughout the orbit.

The Doppler shifts and broadening of spectral lines also decrease with inclination. This is to be expected for the rotational contribution, but also is seen for the influence of the winds, which are dominated by the eastward equatorial jet and so generally contribute in the same manner as the rotation. There is some flow around and over the poles, but this is much weaker and so has negligible influence at low inclinations.

This work shows the inherent advantages and disadvantages of observing planets in non-edge-on geometries using high-resolution spectroscopy. The Doppler effect size and orbital variation both decrease with decreasing inclination, meaning that these signatures may not be accessible, but also may not need to be included in the analysis to accurately match the data. However, it is still possible that 3-D effects may influence the spectra for near pole-on viewing inclinations. \citet{Beltz} compared using 1- or 3-D models in cross correlation with high-resolution emission spectra of the transiting hot Jupiter HD~209458b and found that the 3-D models achieved a notably higher significance for planet detection. This enhancement was identified to be primarily due to the 3-D models having contributions from multiple thermal components from the visible hemisphere of the planet. When viewing a hemisphere of the planet more aligned with the pole, we should expect to see contributions from hotter and cooler regions (e.g., Figures \ref{fig:hemis} and \ref{fig:incs}) and so we may likewise find that multiple thermal components are detectable in high-resolution emission spectra of non-transiting planets, depending on the sensitivity of the data.

In the upcoming decade, new high-resolution spectrographs on 30-m class telescopes will allow for the characterization and analysis of spectra from many times more exoplanets, and to a much higher level of detail than currently available. The Thirty Meter Telescope (TMT), Giant Magellan Telescope (GMT), and European Extremely Large Telescope (E-ELT) all have instruments and science cases designed to be able to revolutionize our characterization of exoplanet atmospheres. While the presence of clouds is already detectable in high resolution spectra with current generation instruments in transmission \citep{Pino2018, Hood2020, Ghandi2020}, we should also be able to see their more subtle influence, if present, in the retrieved temperature profiles of emission spectra (e.g., Line et al., in press). The slight increase in planet detection that \cite{Beltz} found when including Doppler effects in their simulated spectra implies that the stronger Doppler shifts induced in partially cloudy spectra may similarly be at the end of detection limits with current instrumentation. It will be interesting for future work to confront data with simulated spectra for clear and cloudy models, to test whether these signatures already exist in extant data. Within this landscape, our capability for sophisticated modeling of simulated spectra from non-transiting planets can help to interpret current high-resolution spectroscopic observations and guide future observational efforts.

\acknowledgements
IM would like to thank colleagues in his graduate program and Marianne Cowherd, who provided editorial suggestions on drafts of this manuscript. IM would also like to thank Jisheng Zhang, who provided support adapting the radiative transfer code. This research was supported by NASA Astrophysics Theory Program grant NNX17AG25G and the Heising-Simons Foundation. C.K.H. acknowledges support from the National Science Foundation Graduate Research Fellowship Program under Grant No. DGE1752814.

\bibliography{bib}{}

\begin{thebibliography}{}
\expandafter\ifx\csname natexlab\endcsname\relax\def\natexlab#1{#1}\fi
\providecommand{\url}[1]{\href{#1}{#1}}
\providecommand{\dodoi}[1]{doi:~\href{http://doi.org/#1}{\nolinkurl{#1}}}
\providecommand{\doeprint}[1]{\href{http://ascl.net/#1}{\nolinkurl{http://ascl.net/#1}}}
\providecommand{\doarXiv}[1]{\href{https://arxiv.org/abs/#1}{\nolinkurl{https://arxiv.org/abs/#1}}}

\bibitem[{{Batalha} {et~al.}(2019){Batalha}, {Marley}, {Lewis}, \&
  {Fortney}}]{Batalha2019}
{Batalha}, N.~E., {Marley}, M.~S., {Lewis}, N.~K., \& {Fortney}, J.~J. 2019,
  \apj, 878, 70, \dodoi{10.3847/1538-4357/ab1b51}

\bibitem[{{Baxter} {et~al.}(2020){Baxter}, {D{\'e}sert}, {Parmentier}, {Line},
  {Fortney}, {Arcangeli}, {Bean}, {Todorov}, \&
  {Mansfield}}]{2020A&A...639A..36B}
{Baxter}, C., {D{\'e}sert}, J.-M., {Parmentier}, V., {et~al.} 2020, \aap, 639,
  A36, \dodoi{10.1051/0004-6361/201937394}

\bibitem[{{Beatty} {et~al.}(2019){Beatty}, {Marley}, {Gaudi}, {Col{\'o}n},
  {Fortney}, \& {Showman}}]{2019AJ....158..166B}
{Beatty}, T.~G., {Marley}, M.~S., {Gaudi}, B.~S., {et~al.} 2019, \aj, 158, 166,
  \dodoi{10.3847/1538-3881/ab33fc}

\bibitem[{{Beltz} {et~al.}(2021){Beltz}, {Rauscher}, {Brogi}, \&
  {Kempton}}]{Beltz}
{Beltz}, H., {Rauscher}, E., {Brogi}, M., \& {Kempton}, E. M.~R. 2021, \aj,
  161, 1, \dodoi{10.3847/1538-3881/abb67b}

\bibitem[{{Birkby}(2018)}]{2018arXiv180604617B}
{Birkby}, J.~L. 2018, arXiv e-prints, arXiv:1806.04617.
\newblock \doarXiv{1806.04617}

\bibitem[{Butler {et~al.}(2006)Butler, Wright, Marcy, Fischer, Vogt, Tinney,
  Jones, Carter, Johnson, McCarthy, \& Penny}]{Butler_2006}
Butler, R.~P., Wright, J.~T., Marcy, G.~W., {et~al.} 2006, ApJ, 646, 505,
  \dodoi{10.1086/504701}

\bibitem[{Christopher(1989)}]{Mckay1989}
Christopher. 1989, Icarus, 80, 23,
  \dodoi{https://doi.org/10.1016/0019-1035(89)90160-7}

\bibitem[{{Crossfield} {et~al.}(2013){Crossfield}, {Barman}, {Hansen}, \&
  {Howard}}]{Crossfield2013}
{Crossfield}, I. J.~M., {Barman}, T., {Hansen}, B. M.~S., \& {Howard}, A.~W.
  2013, \aap, 559, A33, \dodoi{10.1051/0004-6361/201322278}

\bibitem[{{Crossfield} {et~al.}(2010){Crossfield}, {Hansen}, {Harrington},
  {Cho}, {Deming}, {Menou}, \& {Seager}}]{2010ApJ...723.1436C}
{Crossfield}, I. J.~M., {Hansen}, B. M.~S., {Harrington}, J., {et~al.} 2010,
  \apj, 723, 1436, \dodoi{10.1088/0004-637X/723/2/1436}

\bibitem[{{de Rooij} \& {van der Stap}(1984)}]{1984A&A...131..237D}
{de Rooij}, W.~A., \& {van der Stap}, C.~C.~A.~H. 1984, \aap, 131, 237

\bibitem[{{Demory} {et~al.}(2013){Demory}, {de Wit}, {Lewis}, {Fortney},
  {Zsom}, {Seager}, {Knutson}, {Heng}, {Madhusudhan}, {Gillon}, {Barclay},
  {Desert}, {Parmentier}, \& {Cowan}}]{Demory2013}
{Demory}, B.-O., {de Wit}, J., {Lewis}, N., {et~al.} 2013, \apjl, 776, L25,
  \dodoi{10.1088/2041-8205/776/2/L25}

\bibitem[{{Dobbs-Dixon} {et~al.}(2010){Dobbs-Dixon}, {Cumming}, \&
  {Lin}}]{2010ApJ...710.1395D}
{Dobbs-Dixon}, I., {Cumming}, A., \& {Lin}, D.~N.~C. 2010, \apj, 710, 1395,
  \dodoi{10.1088/0004-637X/710/2/1395}

\bibitem[{Flowers {et~al.}(2019)Flowers, Brogi, Rauscher, Kempton, \&
  Chiavassa}]{Flowers_2019}
Flowers, E., Brogi, M., Rauscher, E., Kempton, E. M.-R., \& Chiavassa, A. 2019,
  AJ, 157, 209, \dodoi{10.3847/1538-3881/ab164c}

\bibitem[{{Fraine} {et~al.}(2014){Fraine}, {Deming}, {Benneke}, {Knutson},
  {Jord{\'a}n}, {Espinoza}, {Madhusudhan}, {Wilkins}, \&
  {Todorov}}]{2014Natur.513..526F}
{Fraine}, J., {Deming}, D., {Benneke}, B., {et~al.} 2014, \nat, 513, 526,
  \dodoi{10.1038/nature13785}

\bibitem[{{Gandhi} {et~al.}(2020){Gandhi}, {Brogi}, \& {Webb}}]{Ghandi2020}
{Gandhi}, S., {Brogi}, M., \& {Webb}, R.~K. 2020, \mnras, 498, 194,
  \dodoi{10.1093/mnras/staa2424}

\bibitem[{{Garhart} {et~al.}(2020){Garhart}, {Deming}, {Mandell}, {Knutson},
  {Wallack}, {Burrows}, {Fortney}, {Hood}, {Seay}, {Sing}, {Benneke}, {Fraine},
  {Kataria}, {Lewis}, {Madhusudhan}, {McCullough}, {Stevenson}, \&
  {Wakeford}}]{2020AJ....159..137G}
{Garhart}, E., {Deming}, D., {Mandell}, A., {et~al.} 2020, \aj, 159, 137,
  \dodoi{10.3847/1538-3881/ab6cff}

\bibitem[{{Guillot}(2010)}]{Guillot2010}
{Guillot}, T. 2010, \aap, 520, A27, \dodoi{10.1051/0004-6361/200913396}

\bibitem[{{Guillot} {et~al.}(1996){Guillot}, {Burrows}, {Hubbard}, {Lunine}, \&
  {Saumon}}]{1996ApJ...459L..35G}
{Guillot}, T., {Burrows}, A., {Hubbard}, W.~B., {Lunine}, J.~I., \& {Saumon},
  D. 1996, \apjl, 459, L35, \dodoi{10.1086/309935}

\bibitem[{{Harada} {et~al.}(2021){Harada}, {Kempton}, {Rauscher}, {Roman},
  {Malsky}, {Brinjikji}, \& {DiTomasso}}]{harada2021}
{Harada}, C.~K., {Kempton}, E. M.~R., {Rauscher}, E., {et~al.} 2021, \apj, 909,
  85, \dodoi{10.3847/1538-4357/abdc22}

\bibitem[{{Harrington} {et~al.}(2006){Harrington}, {Hansen}, {Luszcz},
  {Seager}, {Deming}, {Menou}, {Cho}, \& {Richardson}}]{2006Sci...314..623H}
{Harrington}, J., {Hansen}, B.~M., {Luszcz}, S.~H., {et~al.} 2006, Science,
  314, 623, \dodoi{10.1126/science.1133904}

\bibitem[{Helling(2019)}]{helling2019exoplanet}
Helling, C. 2019, Annual Review of Earth and Planetary Sciences, 47, 583

\bibitem[{Heng {et~al.}(2011)Heng, Menou, \&
  Phillipps}]{10.1111/j.1365-2966.2011.18315.x}
Heng, K., Menou, K., \& Phillipps, P.~J. 2011, Monthly Notices of the Royal
  Astronomical Society, 413, 2380, \dodoi{10.1111/j.1365-2966.2011.18315.x}

\bibitem[{Heng \& Showman(2015)}]{doi:10.1146/annurev-earth-060614-105146}
Heng, K., \& Showman, A.~P. 2015, Annual Review of Earth and Planetary
  Sciences, 43, 509, \dodoi{10.1146/annurev-earth-060614-105146}

\bibitem[{{Hood} {et~al.}(2020){Hood}, {Fortney}, {Line}, {Martin}, {Morley},
  {Birkby}, {Rustamkulov}, {Lupu}, \& {Freedman}}]{Hood2020}
{Hood}, C.~E., {Fortney}, J.~J., {Line}, M.~R., {et~al.} 2020, \aj, 160, 198,
  \dodoi{10.3847/1538-3881/abb46b}

\bibitem[{{Hubeny} \& {Burrows}(2009)}]{2009IAUS..253..239H}
{Hubeny}, I., \& {Burrows}, A. 2009, in IAU Symposium, Vol. 253, Transiting
  Planets, ed. F.~{Pont}, D.~{Sasselov}, \& M.~J. {Holman}, 239--245,
  \dodoi{10.1017/S1743921308026458}

\bibitem[{{Jiang} \& {Ip}(2001)}]{2001A&A...367..943J}
{Jiang}, I.-G., \& {Ip}, W.-H. 2001, \aap, 367, 943,
  \dodoi{10.1051/0004-6361:20000468}

\bibitem[{{Kaeufl} {et~al.}(2004){Kaeufl}, {Ballester}, {Biereichel},
  {Delabre}, {Donaldson}, {Dorn}, {Fedrigo}, {Finger}, {Fischer}, {Franza},
  {Gojak}, {Huster}, {Jung}, {Lizon}, {Mehrgan}, {Meyer}, {Moorwood}, {Pirard},
  {Paufique}, {Pozna}, {Siebenmorgen}, {Silber}, {Stegmeier}, \&
  {Wegerer}}]{Kaeufl2004}
{Kaeufl}, H.-U., {Ballester}, P., {Biereichel}, P., {et~al.} 2004, in Society
  of Photo-Optical Instrumentation Engineers (SPIE) Conference Series, Vol.
  5492, Ground-based Instrumentation for Astronomy, ed. A.~F.~M. {Moorwood} \&
  M.~{Iye}, 1218--1227, \dodoi{10.1117/12.551480}

\bibitem[{Kasting(1991)}]{Kasting1991}
Kasting, J.~F. 1991, Icarus, 94, 1,
  \dodoi{https://doi.org/10.1016/0019-1035(91)90137-I}

\bibitem[{Kataria {et~al.}(2015)Kataria, Showman, Fortney, Stevenson, Line,
  Kreidberg, Bean, \& D{\'{e}}sert}]{Kataria_2015}
Kataria, T., Showman, A.~P., Fortney, J.~J., {et~al.} 2015, ApJ, 801, 86,
  \dodoi{10.1088/0004-637x/801/2/86}

\bibitem[{{Keating} {et~al.}(2019){Keating}, {Cowan}, \&
  {Dang}}]{2019NatAs...3.1092K}
{Keating}, D., {Cowan}, N.~B., \& {Dang}, L. 2019, Nature Astronomy, 3, 1092,
  \dodoi{10.1038/s41550-019-0859-z}

\bibitem[{Kitzmann \& Heng(2017)}]{kitzmann}
Kitzmann, D., \& Heng, K. 2017, Monthly Notices of the Royal Astronomical
  Society, 475, 94, \dodoi{10.1093/mnras/stx3141}

\bibitem[{{Knutson} {et~al.}(2014{\natexlab{a}}){Knutson}, {Benneke}, {Deming},
  \& {Homeier}}]{2014Natur.505...66K}
{Knutson}, H.~A., {Benneke}, B., {Deming}, D., \& {Homeier}, D.
  2014{\natexlab{a}}, \nat, 505, 66, \dodoi{10.1038/nature12887}

\bibitem[{{Knutson} {et~al.}(2014{\natexlab{b}}){Knutson}, {Dragomir},
  {Kreidberg}, {Kempton}, {McCullough}, {Fortney}, {Bean}, {Gillon}, {Homeier},
  \& {Howard}}]{2014ApJ...794..155K}
{Knutson}, H.~A., {Dragomir}, D., {Kreidberg}, L., {et~al.} 2014{\natexlab{b}},
  \apj, 794, 155, \dodoi{10.1088/0004-637X/794/2/155}

\bibitem[{{Lee} {et~al.}(2016){Lee}, {Dobbs-Dixon}, {Helling}, {Bognar}, \&
  {Woitke}}]{lee+2016}
{Lee}, G., {Dobbs-Dixon}, I., {Helling}, C., {Bognar}, K., \& {Woitke}, P.
  2016, \aap, 594, A48, \dodoi{10.1051/0004-6361/201628606}

\bibitem[{{Lee} {et~al.}(2019){Lee}, {Taylor}, {Grimm}, {Baudino}, {Garland},
  {Irwin}, \& {Wood}}]{2019MNRAS.487.2082L}
{Lee}, G. K.~H., {Taylor}, J., {Grimm}, S.~L., {et~al.} 2019, \mnras, 487,
  2082, \dodoi{10.1093/mnras/stz1418}

\bibitem[{{Lee} {et~al.}(2017){Lee}, {Wood}, {Dobbs-Dixon}, {Rice}, \&
  {Helling}}]{lee+2017}
{Lee}, G.~K.~H., {Wood}, K., {Dobbs-Dixon}, I., {Rice}, A., \& {Helling}, C.
  2017, \aap, 601, A22, \dodoi{10.1051/0004-6361/201629804}

\bibitem[{Lines {et~al.}(2019)Lines, Mayne, Manners, Boutle, Drummond,
  Mikal-Evans, Kohary, \& Sing}]{lines2019overcast}
Lines, S., Mayne, N., Manners, J., {et~al.} 2019, Monthly Notices of the Royal
  Astronomical Society, 488, 1332

\bibitem[{{Lines} {et~al.}(2018){Lines}, {Mayne}, {Boutle}, {Manners}, {Lee},
  {Helling}, {Drummond}, {Amundsen}, {Goyal}, {Acreman}, {Tremblin}, \&
  {Kerslake}}]{lines+2018}
{Lines}, S., {Mayne}, N.~J., {Boutle}, I.~A., {et~al.} 2018, \aap, 615, A97,
  \dodoi{10.1051/0004-6361/201732278}

\bibitem[{Marsh {et~al.}(2007)Marsh, Mar, \& Jaffe}]{Marsh2007}
Marsh, J.~P., Mar, D.~J., \& Jaffe, D.~T. 2007, Appl. Opt., 46, 3400,
  \dodoi{10.1364/AO.46.003400}

\bibitem[{{Mayne} {et~al.}(2014){Mayne}, {Baraffe}, {Acreman}, {Smith},
  {Browning}, {Sk{\r{a}}lid Amundsen}, {Wood}, {Thuburn}, \&
  {Jackson}}]{2014A&A...561A...1M}
{Mayne}, N.~J., {Baraffe}, I., {Acreman}, D.~M., {et~al.} 2014, \aap, 561, A1,
  \dodoi{10.1051/0004-6361/201322174}

\bibitem[{{Mayor} \& {Queloz}(1995)}]{1995Natur.378..355M}
{Mayor}, M., \& {Queloz}, D. 1995, \nat, 378, 355, \dodoi{10.1038/378355a0}

\bibitem[{Mischna {et~al.}(2000)Mischna, Kasting, Pavlov, \&
  Freedman}]{Mischna2000}
Mischna, M., Kasting, J., Pavlov, A., \& Freedman, R. 2000, Icarus, 145 2, 546

\bibitem[{Mishchenko {et~al.}(1999)Mishchenko, Dlugach, Yanovitskij, \&
  Zakharova}]{MISHCHENKO1999409}
Mishchenko, M.~I., Dlugach, J.~M., Yanovitskij, E.~G., \& Zakharova, N.~T.
  1999, Journal of Quantitative Spectroscopy and Radiative Transfer, 63, 409,
  \dodoi{https://doi.org/10.1016/S0022-4073(99)00028-X}

\bibitem[{{Obermeier} {et~al.}(2016){Obermeier}, {Koppenhoefer}, {Saglia},
  {Henning}, {Bender}, {Kodric}, {Deacon}, {Riffeser}, {Burgett}, {Chambers},
  {Draper}, {Flewelling}, {Hodapp}, {Kaiser}, {Kudritzki}, {Magnier},
  {Metcalfe}, {Price}, {Sweeney}, {Wainscoat}, \&
  {Waters}}]{2016A&A...587A..49O}
{Obermeier}, C., {Koppenhoefer}, J., {Saglia}, R.~P., {et~al.} 2016, \aap, 587,
  A49, \dodoi{10.1051/0004-6361/201527633}

\bibitem[{{Parmentier} \& {Crossfield}(2018)}]{parm2018}
{Parmentier}, V., \& {Crossfield}, I. J.~M. 2018, {Exoplanet Phase Curves:
  Observations and Theory}, ed. H.~J. {Deeg} \& J.~A. {Belmonte}, 116,
  \dodoi{10.1007/978-3-319-55333-7\_116}

\bibitem[{Parmentier {et~al.}(2016)Parmentier, Fortney, Showman, Morley, \&
  Marley}]{parmientier2016}
Parmentier, V., Fortney, J., Showman, A., Morley, C., \& Marley, M. 2016, ApJ,
  828, \dodoi{10.3847/0004-637X/828/1/22}

\bibitem[{Parmentier {et~al.}(2021)Parmentier, Showman, \&
  Fortney}]{parmentier2021cloudy}
Parmentier, V., Showman, A.~P., \& Fortney, J.~J. 2021, Monthly Notices of the
  Royal Astronomical Society, 501, 78

\bibitem[{{Parmentier} {et~al.}(2013){Parmentier}, {Showman}, \&
  {Lian}}]{2013A&A...558A..91P}
{Parmentier}, V., {Showman}, A.~P., \& {Lian}, Y. 2013, \aap, 558, A91,
  \dodoi{10.1051/0004-6361/201321132}

\bibitem[{{Pino} {et~al.}(2018){Pino}, {Ehrenreich}, {Allart}, {Lovis},
  {Brogi}, {Malik}, {Nascimbeni}, {Pepe}, \& {Piotto}}]{Pino2018}
{Pino}, L., {Ehrenreich}, D., {Allart}, R., {et~al.} 2018, \aap, 619, A3,
  \dodoi{10.1051/0004-6361/201832986}

\bibitem[{Piskorz {et~al.}(2017)Piskorz, Benneke, Crockett, Lockwood, Blake,
  Barman, Bender, Carr, \& Johnson}]{Piskorz_2017}
Piskorz, D., Benneke, B., Crockett, N.~R., {et~al.} 2017, AJ, 154, 78,
  \dodoi{10.3847/1538-3881/aa7dd8}

\bibitem[{Powell {et~al.}(2018)Powell, Zhang, Gao, \& Parmentier}]{Powell_2018}
Powell, D., Zhang, X., Gao, P., \& Parmentier, V. 2018, ApJ, 860, 18,
  \dodoi{10.3847/1538-4357/aac215}

\bibitem[{Ranjan \& Sasselov(2017)}]{Ranjan2017}
Ranjan, S., \& Sasselov, D.~D. 2017, Astrobiology, 17, 169,
  \dodoi{10.1089/ast.2016.1519}

\bibitem[{{Rasio} {et~al.}(1996){Rasio}, {Tout}, {Lubow}, \&
  {Livio}}]{1996ApJ...470.1187R}
{Rasio}, F.~A., {Tout}, C.~A., {Lubow}, S.~H., \& {Livio}, M. 1996, \apj, 470,
  1187, \dodoi{10.1086/177941}

\bibitem[{Rauscher \& Menou(2010)}]{RauscherMenou2010}
Rauscher, E., \& Menou, K. 2010, ApJ, 714, 1334

\bibitem[{Rauscher \& Menou(2012)}]{rauscher2012general}
---. 2012, ApJ, 750, 96

\bibitem[{{Roman} \& {Rauscher}(2017)}]{RomanRauscher2017}
{Roman}, M., \& {Rauscher}, E. 2017, \apj, 850, 17,
  \dodoi{10.3847/1538-4357/aa8ee4}

\bibitem[{{Roman} \& {Rauscher}(2019)}]{2019ApJ...872....1R}
---. 2019, \apj, 872, 1, \dodoi{10.3847/1538-4357/aafdb5}

\bibitem[{Roman {et~al.}(2021)Roman, Kempton, Rauscher, Harada, Bean, \&
  Stevenson}]{roman2021clouds}
Roman, M.~T., Kempton, E. M.-R., Rauscher, E., {et~al.} 2021, ApJ, 908, 101

\bibitem[{{Santos} {et~al.}(2004){Santos}, {Israelian}, \&
  {Mayor}}]{santos2013}
{Santos}, N.~C., {Israelian}, G., \& {Mayor}, M. 2004, \aap, 415, 1153,
  \dodoi{10.1051/0004-6361:20034469}

\bibitem[{Schwartz \& Cowan(2015)}]{10.1093/mnras/stv470}
Schwartz, J.~C., \& Cowan, N.~B. 2015, Monthly Notices of the Royal
  Astronomical Society, 449, 4192, \dodoi{10.1093/mnras/stv470}

\bibitem[{Seager \& Deming(2010)}]{doi:10.1146/annurev-astro-081309-130837}
Seager, S., \& Deming, D. 2010, Annual Review of Astronomy and Astrophysics,
  48, 631, \dodoi{10.1146/annurev-astro-081309-130837}

\bibitem[{{Showman} {et~al.}(2013){Showman}, {Fortney}, {Lewis}, \&
  {Shabram}}]{2013ApJ...762...24S}
{Showman}, A.~P., {Fortney}, J.~J., {Lewis}, N.~K., \& {Shabram}, M. 2013,
  \apj, 762, 24, \dodoi{10.1088/0004-637X/762/1/24}

\bibitem[{{Showman} {et~al.}(2009){Showman}, {Fortney}, {Lian}, {Marley},
  {Freedman}, {Knutson}, \& {Charbonneau}}]{2009ApJ...699..564S}
{Showman}, A.~P., {Fortney}, J.~J., {Lian}, Y., {et~al.} 2009, \apj, 699, 564,
  \dodoi{10.1088/0004-637X/699/1/564}

\bibitem[{{Showman} \& {Guillot}(2002)}]{2002A&A...385..166S}
{Showman}, A.~P., \& {Guillot}, T. 2002, \aap, 385, 166,
  \dodoi{10.1051/0004-6361:20020101}

\bibitem[{{Snellen} {et~al.}(2010){Snellen}, {de Kok}, {de Mooij}, \&
  {Albrecht}}]{2010Natur.465.1049S}
{Snellen}, I. A.~G., {de Kok}, R.~J., {de Mooij}, E. J.~W., \& {Albrecht}, S.
  2010, \nat, 465, 1049, \dodoi{10.1038/nature09111}

\bibitem[{{Stevenson} {et~al.}(2014){Stevenson}, {D{\'e}sert}, {Line}, {Bean},
  {Fortney}, {Showman}, {Kataria}, {Kreidberg}, {McCullough}, {Henry},
  {Charbonneau}, {Burrows}, {Seager}, {Madhusudhan}, {Williamson}, \&
  {Homeier}}]{2014Sci...346..838S}
{Stevenson}, K.~B., {D{\'e}sert}, J.-M., {Line}, M.~R., {et~al.} 2014, Science,
  346, 838, \dodoi{10.1126/science.1256758}

\bibitem[{Stevenson {et~al.}(2017)Stevenson, Line, Bean, D{\'{e}}sert, Fortney,
  Showman, Kataria, Kreidberg, \& Feng}]{Stevenson_2017}
Stevenson, K.~B., Line, M.~R., Bean, J.~L., {et~al.} 2017, AJ, 153, 68,
  \dodoi{10.3847/1538-3881/153/2/68}

\bibitem[{{Toon} {et~al.}(1989){Toon}, {McKay}, {Ackerman}, \&
  {Santhanam}}]{1989JGR....9416287T}
{Toon}, O.~B., {McKay}, C.~P., {Ackerman}, T.~P., \& {Santhanam}, K. 1989,
  \jgr, 94, 16287, \dodoi{10.1029/JD094iD13p16287}

\bibitem[{{van Belle} \& {von Braun}(2009)}]{2009ApJ...694.1085V}
{van Belle}, G.~T., \& {von Braun}, K. 2009, \apj, 694, 1085,
  \dodoi{10.1088/0004-637X/694/2/1085}

\bibitem[{{Wright} {et~al.}(2012){Wright}, {Marcy}, {Howard}, {Johnson},
  {Morton}, \& {Fischer}}]{Wright2012}
{Wright}, J.~T., {Marcy}, G.~W., {Howard}, A.~W., {et~al.} 2012, \apj, 753,
  160, \dodoi{10.1088/0004-637X/753/2/160}

\bibitem[{{Zhang} {et~al.}(2017){Zhang}, {Kempton}, \&
  {Rauscher}}]{2017ApJ...851...84Z}
{Zhang}, J., {Kempton}, E. M.~R., \& {Rauscher}, E. 2017, \apj, 851, 84,
  \dodoi{10.3847/1538-4357/aa9891}

\end{thebibliography}
\bibliographystyle{aasjournal}

\appendix

A major contribution to the work presented here  is that we upgraded the radiative transfer code we use to calculate simulated emission spectra from the 3-D structure output from the GCM. As stated in \S~\ref{sec:Methods}, we have now implemented the two-stream radiative transfer scheme from \cite{1989JGR....9416287T}, which newly enables us to include the influence of multiple scattering in the radiative transfer. In this Appendix we compare the new and old versions of our code, under idealized and fully 3-D conditions. We also show test cases with prescribed values for the scattering parameters, demonstrating expected behavior in the calculated spectra. Finally, we assess the numerical convergence of the code as a function of the number of layers used in the calculation.

\begin{figure}[!htb]
\begin{center}
\includegraphics[width=0.9\linewidth]{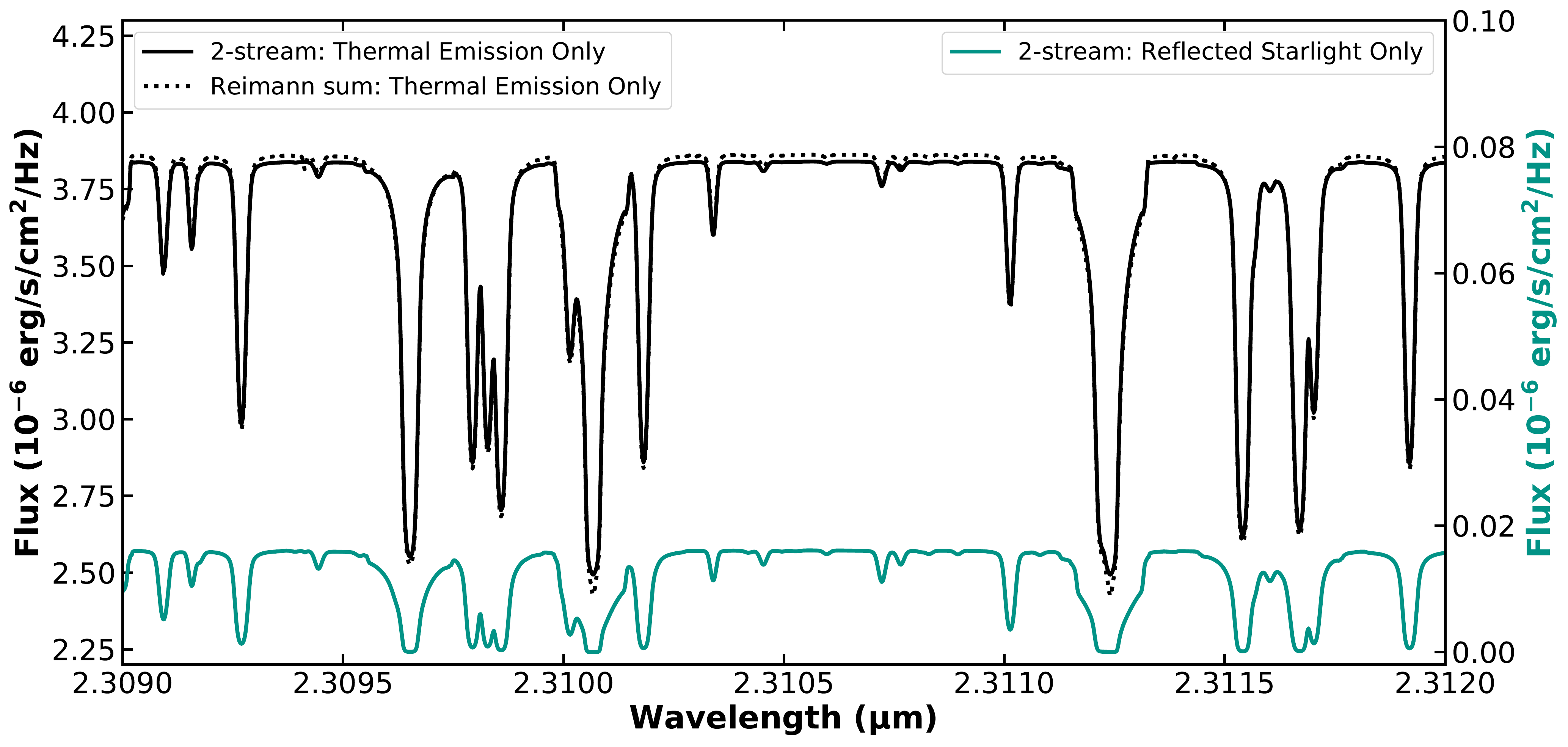}
\caption{Simulated disk integrated emission spectra for the low gravity cloudy model at an orbital inclination of 90$^\circ$ and a phase of 0.5 (so the dayside of the planet is fully in view). The black solid line shows the output flux from only the thermal component of the two-stream radiative transfer scheme. The black dotted line shows the thermal emission only radiative transfer scheme from \cite{2017ApJ...851...84Z}. The teal solid line shows the output flux from only the reflected starlight component of the two-stream radiative transfer scheme. These calculations all used 1000 vertical layers.}
\label{fig:flux_components}
\end{center}
\end{figure}

Figure \ref{fig:flux_components} shows the simulated disk-integrated emission spectra for the individual components of the two-stream solution (thermal emission and reflected starlight), and the previous radiative transfer scheme from \cite{2017ApJ...851...84Z}. The \cite{2017ApJ...851...84Z} code calculates the spectrum in the thermal-emission only limit of the radiative transfer equation, using a Reimann sum integration method.  The thermal components of the two radiative transfer routines are in excellent agreement, to within 0.5\%. Figure \ref{fig:flux_components} also shows the reflected starlight component for the low gravity cloudy model, with an orbital inclination of 90$^\circ$ and at a phase of 0.5. For this planet the reflected starlight makes only a small contribution to the total outgoing flux at these wavelengths; however, a planet with a smaller orbital separation and/or higher albedo could have a more significant reflected component, and our upgraded code is now able to accurately capture that contribution.

\begin{figure}[!htb]
\begin{center}
\includegraphics[width=\linewidth]{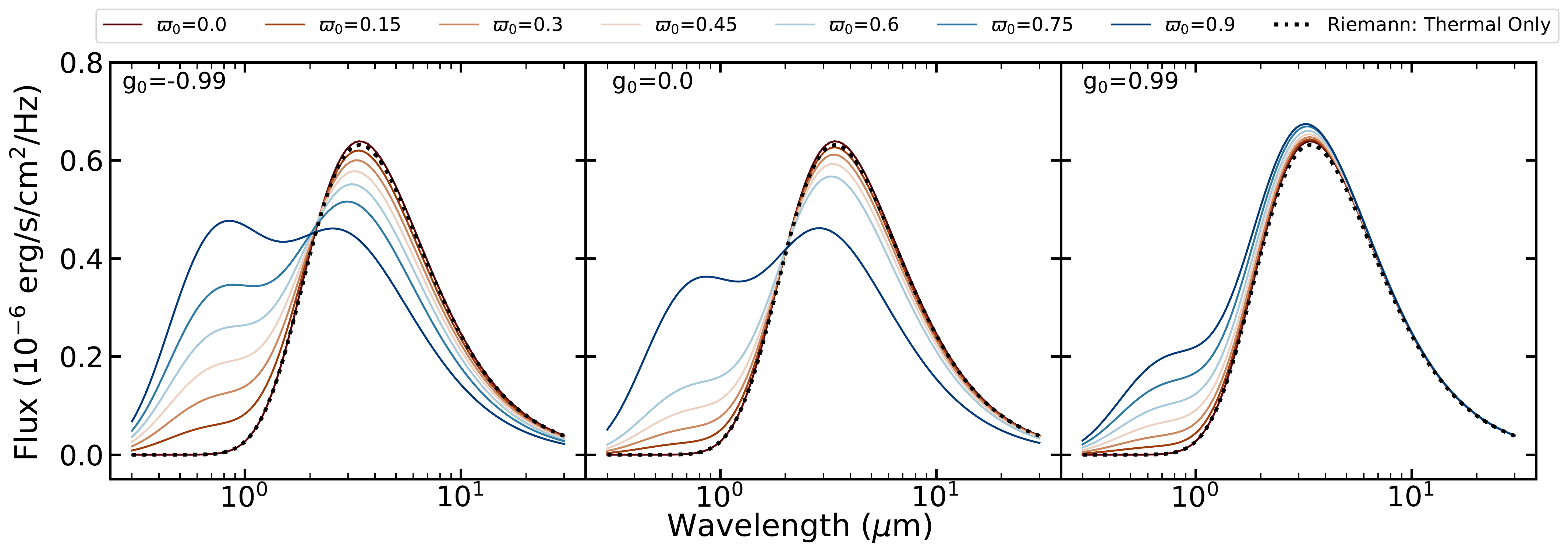}
\caption{Simulated emission spectra for a single isothermal (T=1500 K) line-of-sight column with incident starlight normal to its surface, observed at a phase of 0.5 and calculated using 1000 layers. The left panel shows spectra with g$_0$ = -0.99 (for almost complete back-scattering), the middle panel shows g$_0$ = 0 (isotropic scattering), and the right panel shows  g$_0$ = 0.99 (almost complete forward scattering). All models were simulated with a range of single-scattering albedos from $\varpi_0$ = 0 to 0.9 (none to almost fully reflective). The black dotted line in the right panel shows a calculation of the thermal component using the old version \citep{2017ApJ...851...84Z} of our radiative transfer (which did not include scattering) for comparison, also using 1000 layers.}
\label{fig:2stream-column}
\end{center}
\end{figure}

We use an idealized set-up to test our code's performance over a range of scattering parameters. Figure \ref{fig:2stream-column} shows the thermal and reflected starlight components for a single isothermal column (at T=1500 K) with incident starlight normal to its surface and the observer viewing from the same direction (equivalent to orbital phase 0.5).
We modeled the star with an effective temperature of 6212 K and a semi-major axis of 0.0595 au. We show cases where the single-scattering albedo ($\varpi_0$) and asymmetry parameter (g$_0$) are set to be uniform throughout the column, with values over the ranges indicated on the plot. We also show these results in comparison to the thermal component calculated from the previous version of our code. At shorter wavelengths, the reflected light component increases with increasing $\varpi_0$, as expected. We also see that this component is larger for lower asymmetry parameters, as more of the incident starlight is preferentially scattered back toward the observer at this orientation. Changing the asymmetry parameter also changes the thermal component, but in the opposite sense. Since the direction of net thermal emission is toward the observer, it is forward scattering that preferentially adds scattered light to the observed thermal component, while back-scattering removes flux. For all asymmetry parameters, when $\varpi_0$=0.0 the result reduces to pure thermal emission only, which is also in good agreement with our previous radiative transfer treatment.

\begin{figure}[!htb]
\begin{center}
\includegraphics[width=0.9\linewidth]{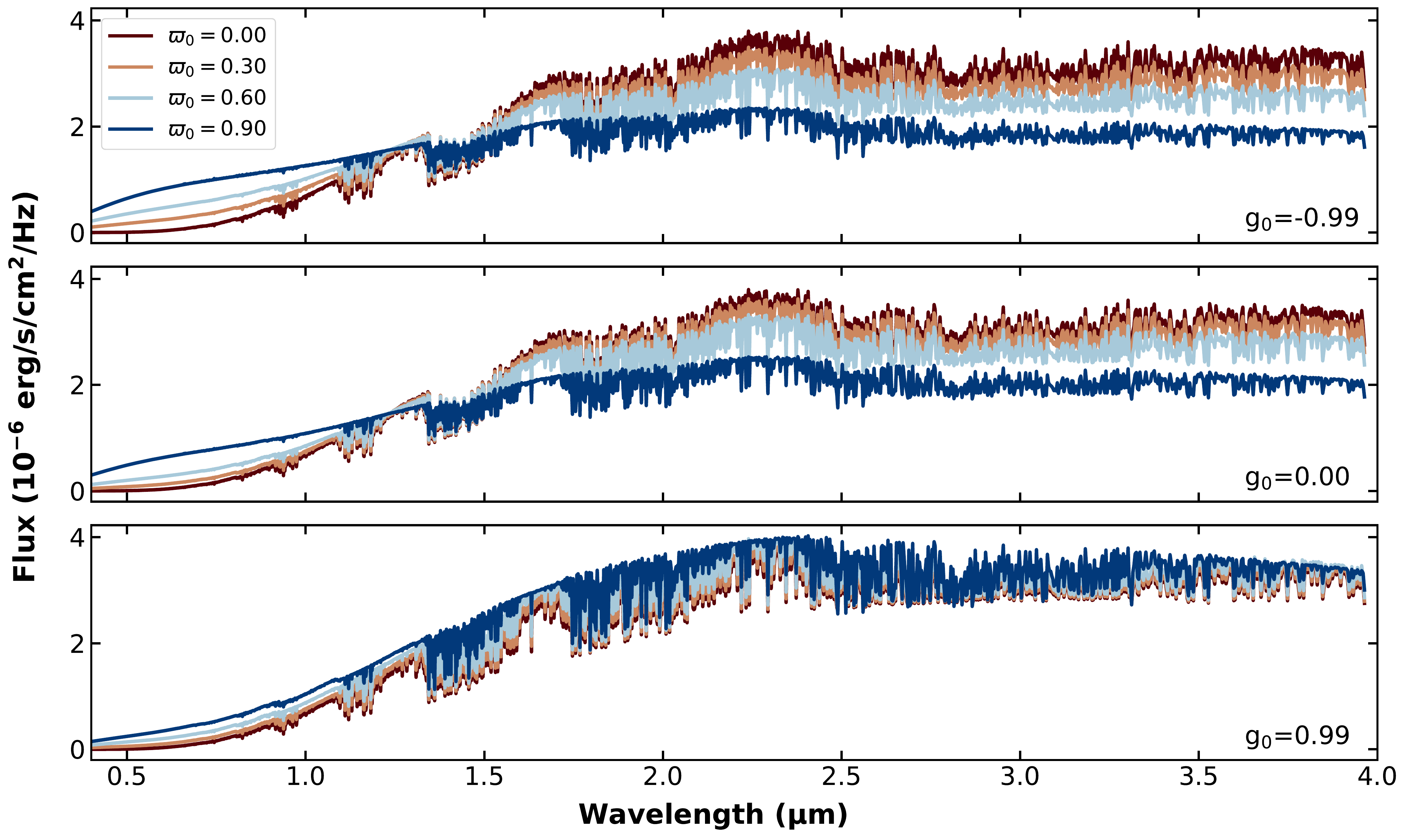}
\caption{Simulated disk integrated emission spectra for the low gravity cloudy model at an orbital inclination of 90$^\circ$ and phase of 0.5, when the dayside of the planet is most aligned with the observer. These models were simulated from the 3-D temperature and cloud structure predicted by our low gravity cloudy GCM, but with artificially overwritten scattering parameters, with uniform single-scattering albedos ranging from 0 to 0.9 (shown as different colors) and asymmetry parameters of g$_0$ = -0.99 (almost complete back-scattering, top panel), g$_0$ = 0 (isotropic scattering, middle panel), and g$_0$ = 0.99 (almost complete forward scattering, bottom panel).}
\label{fig:2stream-low-res1}
\end{center}
\end{figure}

\begin{figure}[!htb]
\begin{center}
\includegraphics[width=0.9\linewidth]{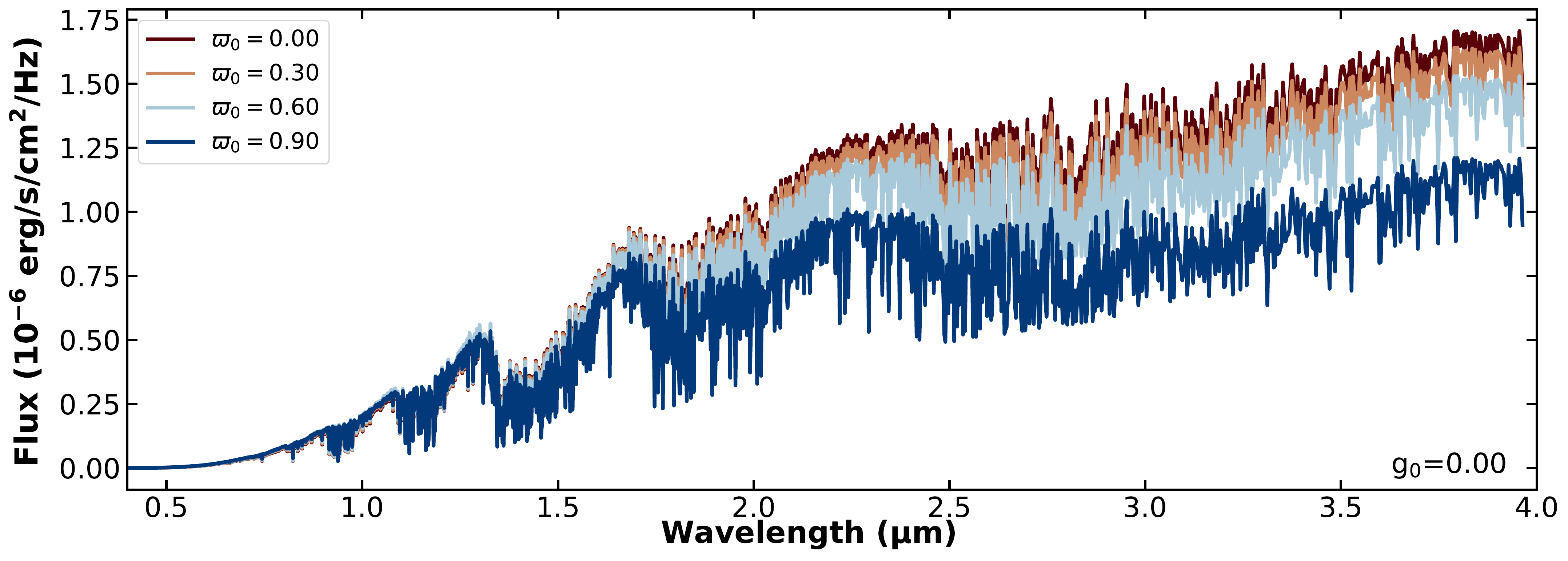}
\caption{Simulated disk integrated emission spectra for the low gravity cloudy model at an orbital inclination of 0$^\circ$ and phase of 0, when the nightside of the planet is aligned with the observer. This is the same model and artificial scattering tests as in Figure \ref{fig:2stream-low-res1}, but with the opposite hemisphere observed.}
\label{fig:2stream-low-res0}
\end{center}
\end{figure}

We then run test cases using a fully 3-D atmospheric structure from one of our models, but with the scattering properties artificially changed as in the idealized isothermal case.
In Figures \ref{fig:2stream-low-res1} and \ref{fig:2stream-low-res0} we show these more complex spectra, over a wider range of wavelengths than plotted in Figure \ref{fig:flux_components} so that we can more clearly see both the reflected and thermal components. Here we use the temperature and cloud structure from our low gravity cloudy model, but overwrite the single-scattering albedo and asymmetry parameter at each location in the atmosphere to prescribed and uniform values, as indicated on the plots. In general, $\varpi_0$ and $g_0$ are wavelength and species dependent, but for testing purposes we set them to uniform values. Figure \ref{fig:2stream-low-res1} shows spectra calculated for the dayside of the planet (at an orbital inclination of 90$^\circ$ and phase of 0.5) while Figure \ref{fig:2stream-low-res0} shows the nightside (90$^\circ$ and 0). We see the same sort of behavior as in Figure \ref{fig:2stream-column}, with the starlight and thermal emission components (at shorter and longer wavelengths, respectively) responding oppositely to the directionality of the scattering. For the nightside spectrum (Figure \ref{fig:2stream-low-res0}), where there is no incident starlight to be reflected, the thermal component is still influenced by internal scattering, in the same manner as we see in the dayside spectrum.

\begin{figure}[!htb]
\begin{center}
\includegraphics[width=0.8\linewidth]{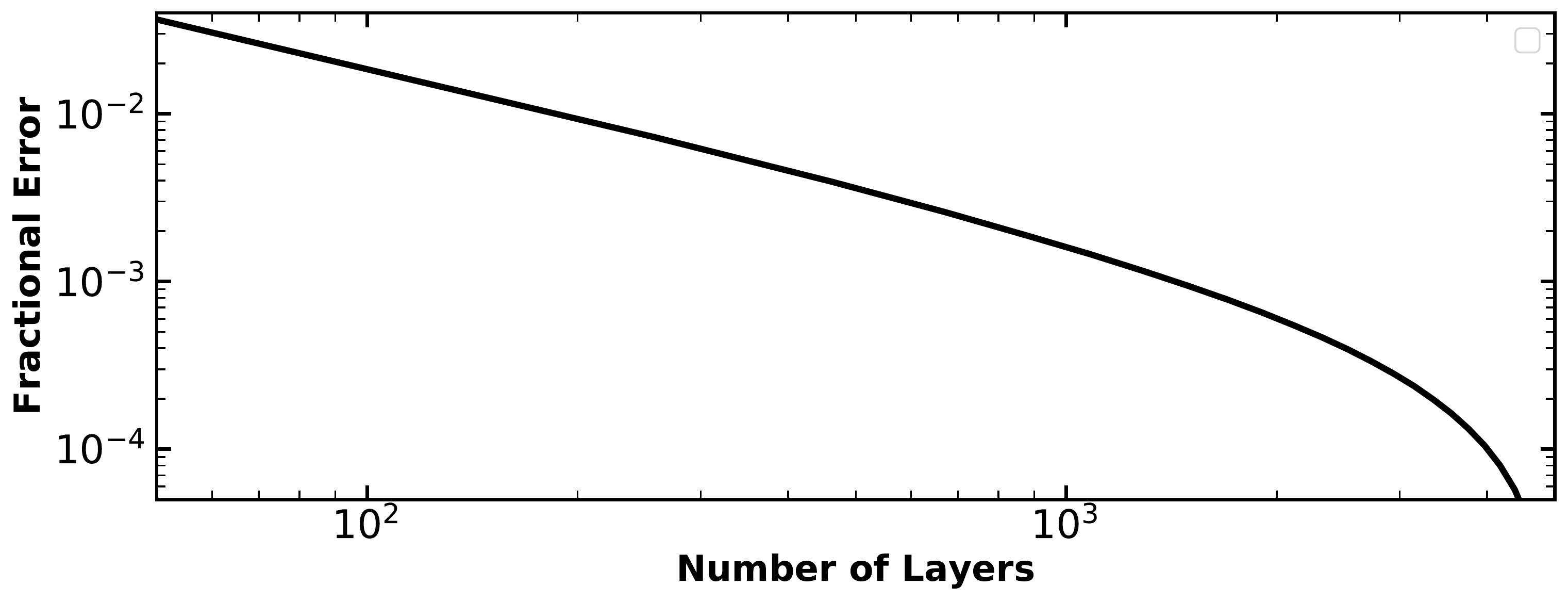}
\caption{The fractional error for the outgoing flux of the two-stream radiative transfer routine. The temperature and optical depth values were taken from a single line-of-sight column of the low gravity clear model at a phase of 0.0 and an inclination of 90$^\circ$, and then interpolated for the varying number of layers. For all simulations we use scattering properties of $\varpi_0$ = 0.0 and $g_0$=0.0. The fractional error is defined in comparison to a calculation from the same model with 5000 layers.}
\label{fig:layer-test}
\end{center}
\end{figure}

Our new two-stream scheme converges to a precise solution with a sufficient number of layers, as shown in Figure \ref{fig:layer-test}. By 200 layers, the fractional error is below 1\%. Increasing the number of layers beyond this slightly increases the accuracy of the radiative transfer results, but is more computationally expensive. We expect other sources of uncertainty in the modeling to dominate at this level of accuracy.

In summary, we have implemented a new two-stream radiative transfer solver into our emission spectrum post-processing code. We have shown that the two-stream calculation behaves as it is expected to in a variety of limiting-case scenarios. Furthermore, we have carefully benchmarked our new code against previous results with no scattering (i.e. \cite{harada2021}, which performed radiative transfer in the pure thermal emission limit), and with scattering turned on (i.e. we find that we can exactly reproduce the visible and IR fluxes from the \cite{RomanRauscher2017} GCM calculations when we run our two-stream code in the double-gray limit). Finally, our two-stream implementation in C produces an exact match to the original \cite{1989JGR....9416287T} routines written in Fortran, which themselves have also been widely incorporated into a variety of other commonly used exoplanet atmosphere models (\citep[e.g.,][]{Mckay1989, Kasting1991, Mischna2000, Ranjan2017, Batalha2019}). As a result of this careful benchmarking exercise, we are confident in the ability of our two-stream post-processing routine to accurately capture the scattered and emitted light from a diverse set of planetary atmospheres both with and without clouds.

\end{document}